\documentclass[10pt,aps,prc,floatfix,twocolumn,nofootinbib, superscriptaddress,preprintnumbers]{revtex4-1}
\usepackage{amsfonts,amsmath,amssymb,bm, xspace}
\usepackage{color,graphicx}
\usepackage{bbm}
\usepackage{dcolumn}                 
\graphicspath{{./figures/}}

\usepackage{hyperref}                

\usepackage{array,physics}
\usepackage{dcolumn}
\usepackage[normalem]{ulem}
\newcolumntype{d}[1]{D{.}{.}{#1}}

\newcommand{\etal}{\textit{et~al.}\xspace}

\newcommand{\kf}{\, k_\textrm{F}}

\newcommand{\fmi}{\, \text{fm}^{-1}}
\newcommand{\fmiq}{\, \text{fm}^{-3}}

\newcommand{\MeV}{\, \text{MeV}}

\newcommand{\nsat}{n_\textrm{sat}}

\newcommand{\NNLO}{\ensuremath{{\rm N}{}^2{\rm LO}}\xspace}
\newcommand{\NNNLO}{\ensuremath{{\rm N}{}^3{\rm LO}}\xspace}

\newcommand{\lsqfit}{\textsl{lsqfit}}

\hypersetup{%
    pdfsubject=Paper,
    pdfkeywords={nuclear physics} {MBPT} {chiral EFT} {asymmetric nuclear matter}
    unicode=true,
    breaklinks=true,
     colorlinks   = true,
     linkcolor = blue,
     citecolor = blue,
     menucolor = blue,
     urlcolor = blue
}

\begin{document}
\preprint{LA-UR-20-26894}
\title{Constraints on the nuclear symmetry energy from asymmetric-matter calculations with chiral NN and 3N interactions}

\author{R. Somasundaram}
\email{somasundaram@ip2i.in2p3.fr}
\affiliation{Univ Lyon, Univ Claude Bernard Lyon 1, CNRS/IN2P3, IP2I Lyon, UMR 5822, F-69622, Villeurbanne, France}
 
\author{C. Drischler}
\email{cdrischler@berkeley.edu}
\affiliation{Department of Physics, University of California, Berkeley, CA 94720, USA}
\affiliation{Nuclear Science Division, Lawrence Berkeley National Laboratory,
 Berkeley, CA 94720, USA}

\author{I. Tews}
\email{itews@lanl.gov}
\affiliation{Theoretical Division, Los Alamos National Laboratory, Los Alamos, New Mexico 87545, USA}

\author{J. Margueron}
\email{j.margueron@ip2i.in2p3.fr}
\affiliation{Univ Lyon, Univ Claude Bernard Lyon 1, CNRS/IN2P3, IP2I Lyon, UMR 5822, F-69622, Villeurbanne, France}

\date{\today}

\begin{abstract}
The nuclear symmetry energy is a key quantity in nuclear (astro)physics. It describes the isospin dependence of the nuclear equation of state (EOS), which is commonly assumed to be almost quadratic.
In this work, we confront this standard quadratic expansion of the EOS with explicit asymmetric nuclear-matter calculations based on a set of commonly used Hamiltonians including two- and three-nucleon forces derived from chiral effective field theory.
We study, in particular, the importance of non-quadratic contributions to the symmetry energy, including the non-analytic logarithmic term introduced by Kaiser [Phys.~Rev.~C \textbf{91}, 065201 (2015)].
Our results suggest that the quartic contribution to the symmetry energy can be robustly determined from the various Hamiltonians employed, and we obtain 1.00(8) MeV (or 0.55(8) MeV for the potential part) at saturation density, while the logarithmic contribution to the symmetry energy is relatively small and model-dependent.
We finally employ the meta-model approach to study the impact of the higher-order contributions on the neutron-star crust-core transition density, and find a small 5\% correction. 
\end{abstract}

\maketitle

\section{Introduction}
\label{sec:introduction}

The nuclear-matter equation of state (EOS) is of great interest for nuclear physics,
see recent reviews~\cite{Oertel2017,FiorellaBurgio2018} and references therein.
It connects bulk properties of atomic nuclei, with small isospin asymmetry, with neutron-rich matter inside neutron stars (NSs)~\cite{NRAPR, Baldo2016}. 
The isospin dependence of the nuclear-matter EOS is described by the nuclear symmetry energy which, for example, governs the proton fraction in beta-equilibrium, determines the pressure in the core of NSs, and hence, the NS mass-radius relation~\cite{Hebeler:2010jx, Gandolfi:2011xu, Steiner:2011ft}, or cooling via the direct URCA process~\cite{Horowitz2014}.
Due to its importance for many physical systems, the symmetry energy and its density dependence were identified as key quantities for nuclear (astro)physics in the 2015 DOE/NSF Nuclear Science Advisory Committee Long Range Plan for Nuclear Science~\cite{LRPNS:2015}, and are actively investigated by combining information from nuclear theory, astrophysics, and experiments.

Because NS observations still come with sizable uncertainties, the symmetry energy and its density dependence can not be inferred from NS properties alone~\cite{Horowitz2014}. 
Hence, various constraints on the symmetry energy have be been inferred from experimental data, e.g., precise determinations of neutron skins in lead (PREX) and calcium~(CREX)~\cite{Abrahamyan2012,Horo14CREXPREX}, collective modes such as giant dipole resonances~\cite{RocaMaza2013}, and heavy-ion collisions~\cite{Tsang:2012se, Lattimer:2014sga}. 
Typically, experimental constraints are in the range $e_{\rm sym}(\nsat) \sim 29-35 \MeV$~\cite{Lattimer:2012xj, Horowitz2014, Margueron2018a} at nuclear saturation density, $\nsat=0.16 \fmiq\equiv \nsat^\mathrm{emp}$ (for a review see, e.g., Ref.~\cite{Tsang:2012se}). 
The determination of the symmetry energy is on the road-map for several future experiments conducted at rare-isotope beam facilities such as FRIB at MSU, SPIRAL2 at GANIL, and FAIR at GSI.
Theoretical calculations provide additional information on the symmetry energy.
While there are nuclear EOS models for a wide range of values for the symmetry energy and its density dependence~\cite{Dutra:2012mb, Dutra:2014qga}, microscopic EOS calculations have improved theoretical constraints over the last years~\cite{Hebeler:2010jx, Gandolfi:2011xu, Steiner:2011ft, Tews:2012fj, Drischler:2015eba, Lonardoni:2019ypg, Drischler:2020a}.

A consistent extraction of the nuclear symmetry energy from nuclear theory as well as experimental and astrophysical programs requires that the measured quantities in these different approaches, as well as their relations, are well defined. 
Different approximations for the symmetry energy are commonly used.
It is, therefore, important to clarify whether the symmetry energy measured in laboratory experiments is the same quantity as the one inferred from NS properties. 
For example, the energy per particle of nuclear matter at zero temperature is a function of the baryon density $n=n_n+n_p$ and isospin asymmetry $\delta=(n_n-n_p)/n$, where $n_n$ ($n_p$) denotes the neutron (proton) number density.
The following isospin-asymmetry expansion from symmetric nuclear matter (SNM, $\delta = 0$) to pure neutron matter (PNM, $\delta = 1$) is often employed,
\begin{eqnarray} \label{eq:EA_quadratic_expansion}
 e(n, \, \delta) &\approx& e(n,\, \delta = 0) + \delta^2 \, e_\text{\textrm{sym},2}(n) \nonumber \\
 && + \delta^4 \, e_\text{\textrm{sym},4}(n) + \mathcal{O}(\delta^6)\,.
\end{eqnarray}
Here $e_\text{\textrm{sym},2}(n)$ and $e_\text{\textrm{sym},4}(n)$ are the quadratic and quartic contributions to the symmetry energy, respectively.
Given the expansion~\eqref{eq:EA_quadratic_expansion}, the quadratic contribution to the symmetry energy is defined by the second derivative
\begin{equation} \label{eq:esym2}
  e_{\textrm{sym},2}(n) = \frac{1}{2} \frac{\partial^2 e(n,\delta)}{\partial \delta^2}\biggr\rvert_{\delta=0} \, ,
\end{equation}
similar to the empirical Bethe-Weizs{\"a}cker mass formula for finite nuclei. 
Hence, $e_\text{\textrm{sym},2}(n)$ is often referred to as the \emph{symmetry energy}, and used in nuclear experiments.
In practice, however, the more commonly used definition of the symmetry energy is given by the difference between the energy per particle in pure neutron matter (PNM) and SNM,
\begin{equation}\label{eq:esym1}
e_{\textrm{sym}}(n) = e_{\textrm{PNM}}(n)-e_{\textrm{SNM}}(n)\,.
\end{equation}
While this definition needs information on the EOS only in the limits of PNM and SNM, Eq.~\eqref{eq:esym2} necessitates explicit calculations of isospin-asymmetric nuclear matter (ANM).
Both $e_{\textrm{sym},2}(n)$ and $e_{\textrm{sym}}(n)$ are equal if the isospin dependence of the energy per particle is \emph{exactly} quadratic, i.e., non-quadratic terms in the expansion~\eqref{eq:EA_quadratic_expansion} vanish.
However, there is no \emph{a priori} argument why these terms should vanish.
In fact, they have been found to be relevant for, e.g., accurate studies of nuclear matter in beta-equilibrium at supra-saturation density~\cite{Steiner2006, Chen2009, Cai2012, Cai2015} and the crust-core transition density in NSs~\cite{Cai2012, Seif2014}.

In this work, we study the expansion~\eqref{eq:EA_quadratic_expansion} and quantify the impact of non-quadratic contributions to the symmetry energy.
We investigate to which extent uncertainties in the microscopical approach affect the extraction of non-quadratic contributions to the symmetry energy.
The paper is organized as follows. 
In Sec.~\ref{sec:previous}, we discuss some previous studies of non-quadratic contributions to the symmetry energy.
In Sec.~\ref{sec:Micro}, we present our computational setup and summarize the results calculated in Ref.~\cite{Drischler:2015eba}.
General expressions for the energy expansion in terms of the isospin-asymmetry parameter $\delta$ are given in Sec.~\ref{sec:expansion}, in particular, for the total energy per particle as well as for the contributions of the potential energy terms. 
We then discuss the EOS in the limits of SNM and PNM in Sec.~\ref{sec:SNMPNM}, followed by the symmetry energy in Sec.~\ref{sec:ESYM}.
In Sec.~\ref{sec:SPINO}, we study the impact of the non-quadratic contributions to the symmetry energy in determinations of the core-crust transition in NSs. 
Finally, we conclude in Sec.~\ref{sec:CONCLUSIONS}.
The Python codes used to perform the analysis and generate the figures in this paper are publicly available on GitHub~\cite{rahul_2020_4010747} and briefly described in the Supplemental Material associated to this publication.

\section{Previous studies of non-quadratic contributions}
\label{sec:previous}

As stated above, there is no \emph{a priori} reason for the isospin-asymmetry expansion to be purely quadratic. 
In general, even the free Fermi gas (FFG) energy per particle, given by
\begin{equation}\label{eq:FFG}
	e^\text{FFG}(n) = \frac{t_{\textrm{SNM}}^{\textrm{sat}}}{2}\left(\frac{n}{n_{\textrm{sat}}}\right)^{2/3} \Big[ (1+\delta)^{5/3}+(1-\delta)^{5/3} \Big] \, ,
\end{equation}
with $t_{\textrm{SNM}}^{\textrm{sat}}= \frac{3}{5m_N}\left(\frac{3\pi^2}{2} n_{\textrm{sat}}\right)^{2/3}\approx 22.1\MeV$, leads to non-quadratic contributions to the expansion~\eqref{eq:EA_quadratic_expansion}.
For example, the quartic term,
\begin{equation}
	e_{\textrm{sym},4}^\text{FFG}(n) \simeq 0.45 \MeV \times \left(\frac{n}{\nsat}\right)^{2/3} \,,
\end{equation}
represents a $\sim3.5$\% correction to the FFG symmetry energy at $\nsat$.
Nuclear interactions also contribute to non-quadratic terms.
For example, the phenomenological Skyrme interaction~\cite{Bender2003} gives the following quartic contribution to the symmetry energy:
\begin{multline}
e_{\textrm{sym},4}^\mathrm{Skyrme}(n) \simeq e_{\textrm{sym},4}^\text{FFG}(n) \\
+\frac{\kf^5}{972\pi^2} \left[3t_1(1+x_1)+ t_2(1-x_2)\right] \, ,
\end{multline}
where the Skyrme parameters ($t_1$, $t_2$) represent the correction to the bare nucleon mass generated by in-medium effects.
Since the Skyrme in-medium mass is generally $\sim 30-40\%$ lower than the bare mass~\cite{Bender2003}, these terms contribute to an increase of the FFG $e_{\textrm{sym},4}$ of $\sim 30-40\%$ to $\sim (0.7-0.8)\MeV$.
In a recent work, Cai and Li~\cite{Cai2015} found $e_{\textrm{sym},4}(n_{\rm sat})=(7.2\pm2.5)$~MeV, which indicates a rather significant difference between $e_{\textrm{sym}}$ and $e_{\textrm{sym},2}$.
They employed an empirically constrained isospin-dependent single-nucleon momentum distribution and the EOS of PNM near the unitary limit. 
Subsequently, Bulgac~\etal found that  $e_{\textrm{sym},4}(n=0.1 \fmiq)=2.635$~MeV is necessary in order to reproduce properties of both finite nuclei and the PNM EOS as calculated in Ref.~\cite{Wlazlowski2014}.
In contrast, previous works, e.g., based on Brueckner-Hartree-Fock (BHF) approaches and hard-core interactions~\cite{Bombaci1991, Lee1998, Frick2005, Vidana2009} obtained only small non-quadratic contributions to the symmetry energy.

In a recent study of nuclear matter in many-body perturbation theory (MBPT) with two-nucleon (NN) and three-nucleon (3N) interactions derived from chiral effective field theory (EFT), Kaiser~\cite{Kaiser2015} did not confirm such large values for $e_{\textrm{sym},4}$.
Instead, Kaiser found $e_{\textrm{sym},4}\simeq 1.5$~MeV at $\nsat$, which is still about three times larger than the FFG contribution.
Moreover, Kaiser found contributions to the energy per particle whose fourth derivative with respect to $\delta$ was singular at $\delta=0$.
This was further substantiated by analytic MBPT calculations based on an $S$-wave contact interaction, which gave rise to a singular term $\propto \delta^4\log\vert\delta\vert$---a term that only contributes to ANM when $\delta\ne 0$ and $\delta\ne 1$, and which will be referred to as the leading-order logarithmic term in the following. 

Subsequently, Wellenhofer~\etal performed a more detailed analysis of such divergences by examining the $\delta$ dependence of the nuclear EOS as a function of density and temperature~\cite{Wellenhofer2016}.
They found that the asymmetry expansion is hierarchically ordered, i.e., the lower-order coefficients are dominant at finite temperature and low density, but the expansion diverges with alternating sign in the zero-temperature limit.
Around saturation density, their results indicate that the convergence of the series expansion is restored for $T\gtrsim 3$~MeV.
Moreover, they have argued that the logarithmic term at leading order considerably improves the isospin-asymmetry expansion at zero temperature and suggested to include this term in future fits of the EOS.

While mathematically well-defined, it is not clear that the aforementioned divergence of the series expansion in the isospin asymmetry parameter substantially impacts the practical usability of the expansion~\eqref{eq:EA_quadratic_expansion}, because corrections remain small at nuclear densities.
As remarked by Wellenhofer\etal, the lack of precision in nonperturbative approaches to the nuclear many-body problem makes it practically impossible to precisely extract the high-order isospin-asymmetry derivatives.
Furthermore, our knowledge of the symmetry energy, and even more fundamentally of the nuclear interaction itself, is limited by experimental precision and by a limited theoretical understanding of strongly-interacting systems.
As a consequence, while the series expansion in the isospin asymmetry can be determined with high accuracy when the nuclear interaction and the many-body treatment are fixed (with numerical limitations as discussed in Ref.~\cite{Wellenhofer2016}),
current theoretical uncertainties reduce our ability to accurately determine high-order contributions in general. In this paper, we analyze the impact of these uncertainties on the determination of the symmetry energy.

\section{Nuclear-matter equation of state}
\label{sec:Micro}

\begin{table}[tb]
    \centering
    \caption{
    Nonlocal \NNNLO NN and \NNLO 3N interactions used in the MBPT calculations
    of Ref.~\cite{Drischler:2015eba}. 
    Each Hamiltonian in the Table is based on the \NNNLO NN
    potential EM~$500\MeV$~\cite{Entem2003} evolved to the SRG resolution scale~$\lambda$.
    The low-energy couplings $c_D$ and $c_E$ were subsequently fit to the triton binding energy and the charge radius of ${}^4\textrm{He}$ in Ref.~\cite{Hebeler2011} for different combinations of $\lambda$ and the 3N cutoff $\Lambda_\textrm{3N}$. 
    The 3N two-pion exchange is governed by the $\pi$N low-energy couplings $c_{1}$, $c_{3}$, and $c_{4}$, which were taken from the NN potential, except for H7 which uses the values obtained from the NN partial-wave analysis (PWA) of Ref.~\cite{Rentmeester2003}. 
    Hamiltonian H6 has been excluded as discussed in Section IV~B of Ref.~\cite{Drischler:2015eba}.
}
    \label{tab:used_hamiltonians}
    \begin{ruledtabular}
    \begin{tabular}{ccclcc}
        label &	$\lambda$ [$\fmi$] &	$\Lambda_\textrm{3N}$ [$\fmi$]& 3N $c_{1,3,4}$ & $c_D$ & $c_E$  \\
        \hline
        H1 & $1.8$ & $2.0$ & NN potential & $+1.264$  & $-0.120$\\
        H2 & $2.0$ & $2.0$ & NN potential & $+1.271$  & $-0.131$\\
        H3 & $2.0$ & $2.5$ & NN potential & $-0.292$  & $-0.592$\\
        H4 & $2.2$ & $2.0$ & NN potential & $+1.214$  & $-0.137$\\
        H5 & $2.8$ & $2.0$ & NN potential & $+1.278$  & $-0.078$\\
        H7 & $2.0$ & $2.0$ & PWA~\cite{Rentmeester2003} & $-3.007$  & $-0.686$
    \end{tabular}
    \end{ruledtabular}
\end{table}

The nuclear-matter EOS has been investigated since the nineteen-fifties employing various theoretical approaches, see e.g. Refs.~\cite{Oertel2017,FiorellaBurgio2018} and references therein. 
Recently, improved studies of nuclear matter and finite nuclei~\cite{Hammer2013, Hebeler2015, Tews:2020hgp} have become available with the advent of chiral EFT~\cite{Epelbaum:2008ga, Machleidt:2011zz} and renormalization group methods~\cite{BOGNER201094, Furnstahl_2013}. 
Importantly, chiral EFT interactions have also enabled theoretical uncertainty estimates~\cite{Epelbaum:2014efa, Drischler:2020a, Drischler:2020b}. 
As for the various many-body approaches, significant progress has been made in predicting the EOS in the limits of PNM and SNM using the self-consistent Green's function method~\cite{Carbone:2013rca, Carbone:2013eqa}, coupled cluster theory~\cite{Hagen:2013yba,Ekstrom2015}, MBPT~\cite{Hebeler2010, Wellenhofer:2014hya, Holt:2016pjb, Drischler:2017wtt, Coraggio2014}, in-medium chiral perturbation theory~\cite{Holt2013}, and Quantum Monte Carlo (QMC) studies~\cite{Gezerlis:2013ipa, Gezerlis:2014zia, Tews2016, Lynn:2015jua, Lonardoni:2019ypg}. 
On the other hand, only a few explicit calculations of ANM exist to date~\cite{Drischler2014, Wellenhofer2015, Kaiser2015, Drischler:2015eba}, because typically ANM is computationally more involved. 

In this work, we use the explicit ANM calculations of Ref.~\cite{Drischler:2015eba} to study the importance of non-quadratic contributions to the symmetry energy. 
We focus our study on the zero-temperature limit in which the dominant effects are expected.
Drischler~\etal evaluated the energy per particle $E/A(n,\delta)$ at 11 isospin asymmetries ($\delta = 0.0, 0.1,\ldots, 1.0$) using MBPT up to second order, and estimated the neglected third-order ladder contributions to be small. 
For each proton fraction, the energy per particle is sampled on an equidistant grid in the neutron Fermi momentum (not the density), and available up to $2 \fmi$ leading to $385$ points in total for each Hamiltonian.
The MBPT calculations of Ref.~\cite{Drischler:2015eba} are based on the set of six Hamiltonians with the chiral NN and 3N interactions summarized in Table~\ref{tab:used_hamiltonians}. These chiral Hamiltonians are also commonly used in nuclear-structure calculations~\cite{Simonis:2015vja,Hage16NatPhys,Ruiz16Calcium,
Hage16Ni78,Simo17SatFinNuc,Birk17dipole,Morr17Tin,Holt:2019gmc,Mougeot:2020mlc,Kaufmann:2020gbf}.
They combine the \NNNLO NN potential EM~$500\MeV$~\cite{Entem2003} evolved to lower momentum scales using the similarity renormalization group (SRG) with bare \NNLO 3N forces regularized by a nonlocal regulator with momentum cutoff $\Lambda_\textrm{3N}$. 
Hebeler~\etal then refit the two 3N low-energy couplings $c_D$ and $c_E$ for different combinations of $\lambda$ and $\Lambda_\textrm{3N}$ shown in Table~\ref{tab:used_hamiltonians} to the triton binding energy as well as the charge radius of ${}^4\text{He}$~\cite{Hebeler2011}. 
Assuming \NNLO 3N forces provide a sufficiently complete operator basis, and the long-range low-energy couplings $c_1$, $c_3$, and $c_4$ are SRG-invariant, this approach captures dominant contributions from induced three- and higher-body forces due to the SRG transformation. Note that the $c_i$'s appear both, in the NN and 3N interactions at \NNLO.
As discussed in Ref.~\cite{Drischler:2015eba}, the spread of energies per particle obtained from these nuclear interactions can serve as a simple uncertainty estimate---although not allowing for a statistical interpretation.

Normal-ordering is the standard approach for including (dominant) 3N contributions in many-body calculations in terms of density-dependent effective two-body potentials. 
In infinite nuclear matter, normal ordering corresponds to summing one particle in the 3N forces over the occupied states of the (e.g., Hartree-Fock) reference state. 
This induces a dependence on the total momentum ${\bf P}$ of the two remaining particles, in contrast to the Galilean invariant NN potentials. 
To straightforwardly combine NN and effective potentials, the approximation ${\bf P} = 0$ had previously been imposed~\cite{Holt10ddnn,Hebeler2010}.
Reference~\cite{Drischler:2015eba} relaxed this approximation by averaging over all directions of $\bf{P}$ using a novel partial-wave-based method that generalizes normal-ordering of arbitrary 3N interactions.
We focus the discussion here on the results obtained with this improved (angle averaging) approximation in a Hartree-Fock single-particle spectrum.

\subsection{Energy per particle}

\begin{figure}
\centering
\includegraphics[scale=0.48]{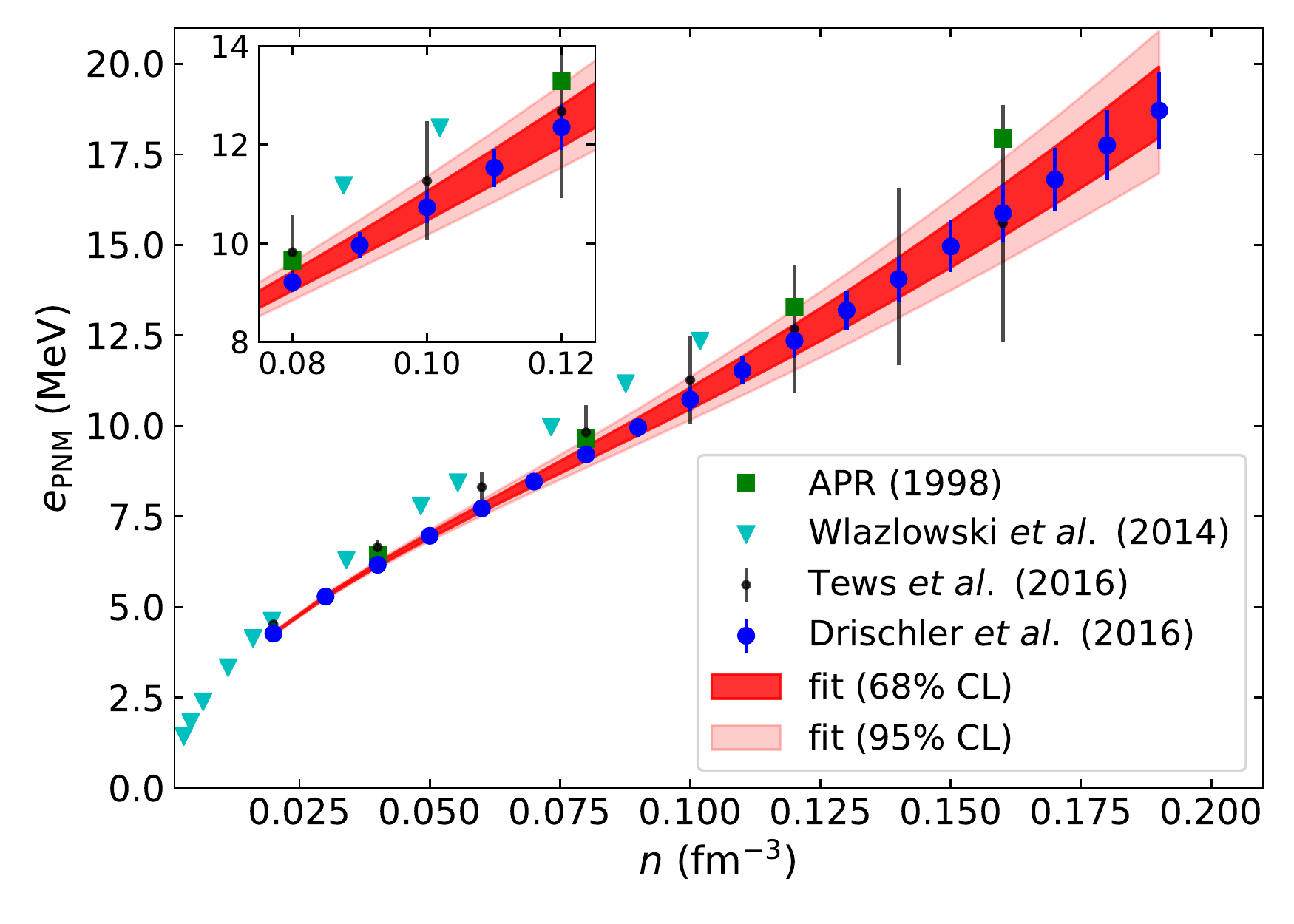}
\includegraphics[scale=0.48]{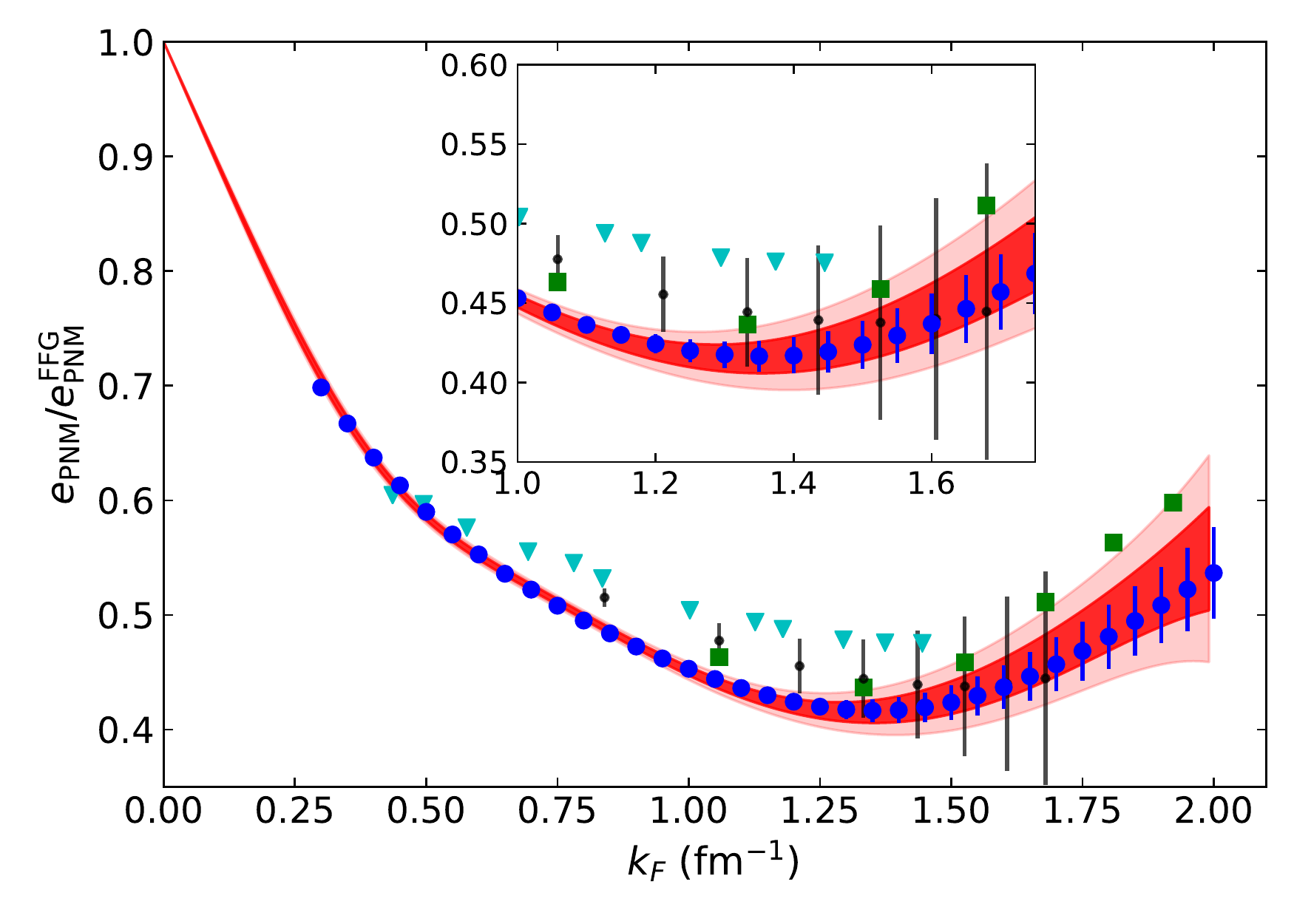}
\caption{Comparison of the MBPT predictions for the energy per particle in PNM~\cite{Drischler:2015eba} (blue points) with the APR EOS~\cite{Akmal1998} (green squares), QMC calculations Wlazlowski \textsl{et al.} (2014)~\cite{Wlazlowski2014} (cyan triangles), and Tews \textsl{et al.} (2016)~\cite{Tews2016} (black dots) using different chiral EFT Hamiltonians with NN and 3N forces.
The latter points include simple estimates for the EFT truncation error of the chiral expansion.
We also show our fit posterior at 68\% (95\%) confidence level as dark (light) red bands. See the main text for details.}
\label{fig:PNM_3_add}
\end{figure}

The energies per particle in PNM predicted by the MBPT approach~\cite{Drischler:2015eba} are shown in the upper panel of Fig.~\ref{fig:PNM_3_add} as blue dots (the lower panel represents the ratio of the energy per particle over the free Fermi gas energy). 
For the fits to the MBPT data that we use later in our analysis, we also show 68\% (95\%) confidence intervals of the fit posteriors as dark (light) red bands. 
In addition, we compare with the variational calculation of Ref.~\cite{Akmal1998} (APR), Fock-space formulated Quantum Monte Carlo (QMC) calculations of Ref.~\cite{Wlazlowski2014} (Wlazlowski \etal~2014), and continuum QMC calculations using auxiliary field diffusion Monte Carlo of Ref.~\cite{Tews2016} (Tews \etal~2016). These calculations were conducted not only using different many-body approaches, but also different nuclear interactions:
the APR result uses the Argonne v18 (AV18) NN potential~\cite{Wiringa:1994wb} and the Urbana IX (UIX) 3N force~\cite{Pudliner:1995wk}, 
Ref.~\cite{Wlazlowski2014} employs the nonlocal momentum-space chiral \NNNLO NN interactions of Ref.~\cite{Coraggio:2007mc} combined with \NNLO 3N forces as specified in Ref.~\cite{Coraggio:2012ca}, 
and Ref.~\cite{Tews2016} uses local coordinate-space chiral interactions constructed in Refs.~\cite{Gezerlis:2013ipa, Gezerlis:2014zia, Lynn:2015jua}.
The first two calculations do not provide theoretical uncertainties, while the latter estimates the standard EFT uncertainty~\cite{Epelbaum:2014efa}. 
Note that, in general, order-by-order calculations are required for estimating EFT truncation errors. Such calculations are not possible with the chiral Hamiltonians given in Table~\ref{tab:used_hamiltonians}.

When comparing the approaches using chiral EFT interactions, the QMC calculations of Ref.~\cite{Tews2016} agree with the MBPT approach employed in this work within uncertainties above $n\sim 0.08 \fmiq$, while QMC finds slightly higher energies at lower densities.
In contrast, the QMC calculations of Ref.~\cite{Wlazlowski2014} find a higher PNM energy per particle at all densities, by about $\sim 1$~MeV.
We also compare the ratio $e_{\textrm{PNM}}/e^\mathrm{FFG}_{\textrm{PNM}}$ as a function of neutron Fermi momentum $\kf$ for the various calculations in the bottom panel of Fig.~\ref{fig:PNM_3_add}. 
In the figure, we can identify the density region where the ratio exhibits a plateau, indicating a similar scaling of $e_{\textrm{PNM}}$ and $e^\text{FFG}$ with $\kf$. 
For the MBPT calculation, we find the ratio at the plateau to be $\approx 0.42(1)$ in PNM at momenta $\kf \approx 1.3(2) \fmi$, which describes densities of $\approx n_{\rm sat}/2$.

The comparison of the different results in Fig.~\ref{fig:PNM_3_add} provides a qualitative illustration of the uncertainties originating from the nuclear interactions as well as from the different many-body approaches.
While the MBPT results of Ref.~\cite{Drischler:2015eba} provide a simple uncertainty estimate, they do not quantify EFT truncation errors, which are the dominant source of theoretical uncertainty.
Strategies to quantify truncation errors have only been employed in the last few years.
Future order-by-order calculations of ANM will enable statistically robust EFT uncertainty estimates using the Bayesian framework recently developed by the BUQEYE collaboration~\cite{Drischler:2020a, Drischler:2020b}.
In the present analysis, however, such calculations are not available.
Therefore, we follow the approach in Ref.~\cite{Drischler:2015eba}, and consider the spread of the EOSs due to the Hamiltonians in Table~\ref{tab:used_hamiltonians} as an uncertainty estimate.

\subsection{Single-particle energies}

\begin{figure*}[t]
    \centering
    \includegraphics[scale=0.50]{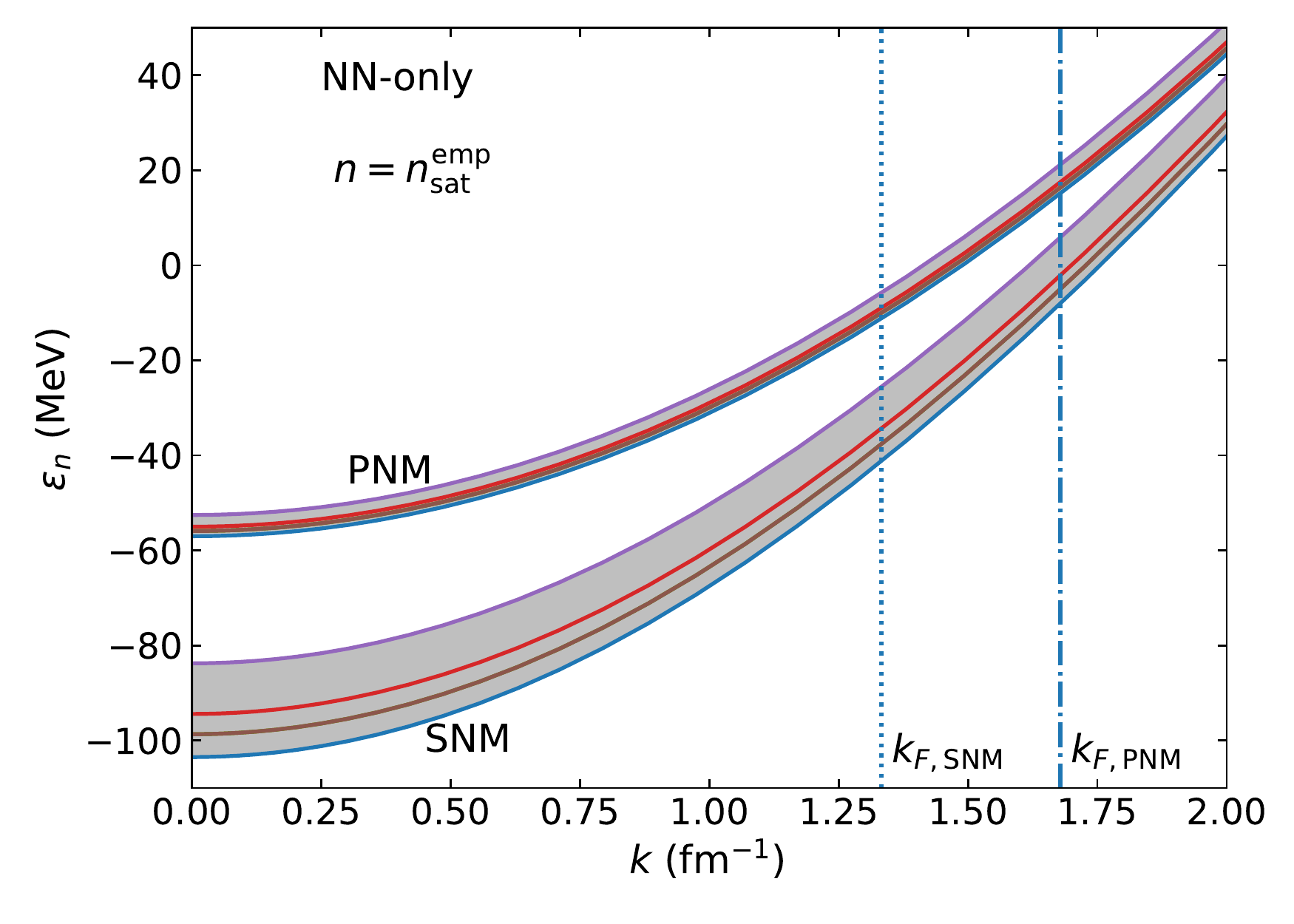}
    \includegraphics[scale=0.50]{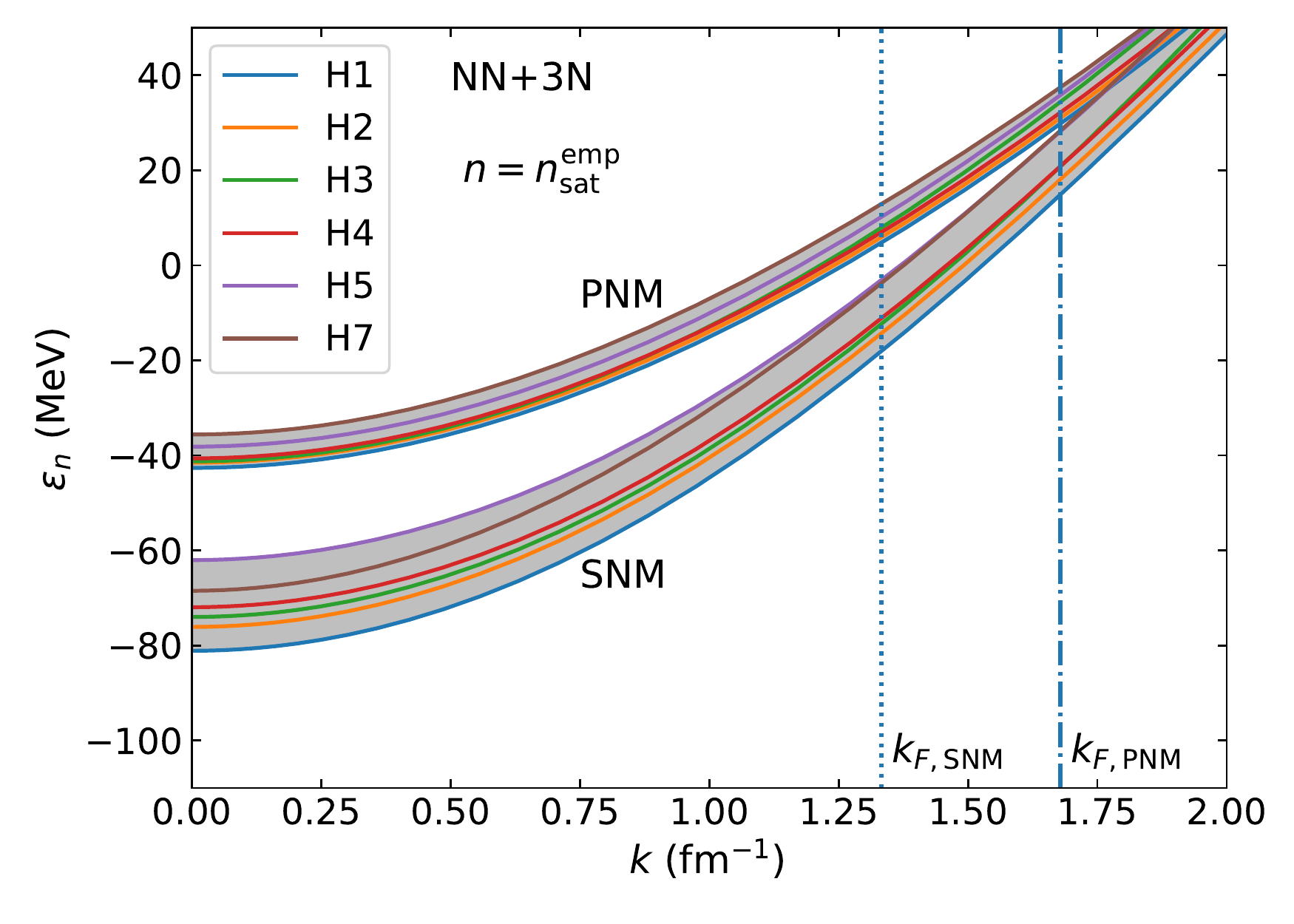}
    \caption{The neutron single-particle energies $\epsilon_n(k)$ as a function of the momentum $k$ calculated at $n_{\textrm{sat}}^\mathrm{emp}$, and extracted from the MBPT calculations of Ref.~\cite{Drischler:2015eba}.
    The different colors correspond to the six Hamiltonians as labeled in the legend.
    We show the single particle energies obtained from only NN forces (left panel) and when including 3N contributions (right panel).
    In each panel, we present results for both SNM and PNM.
    }
    \label{fig:spe}
\end{figure*}

The single particle energy $\epsilon_\tau(k)$ is determined by the self-consistent solution to the Dyson equation. In a Hartree-Fock spectrum it is given by
\begin{equation}
\epsilon_\tau(k,n,\delta)\approx \frac{k^2}{2m_\tau}+\Sigma^{(1)}(k,n,\delta)\,. \label{eq:spe}
\end{equation}
The first term in Eq.~(\ref{eq:spe}) is the single-particle kinetic energy, while the second term $\Sigma^{(1)}$ denotes the spin-isospin-averaged first-order self-energy.
We refer the reader to, e.g., Refs.~\cite{Hebeler2010,Drischler:2015eba} for more details.

Figure~\ref{fig:spe} shows the single-particle energy $\epsilon_n(k)$ in SNM and PNM evaluated at $n_{\textrm{sat}}^\mathrm{emp}$. 
The left (right) panel depicts the NN-only (NN+3N) results, and the vertical lines mark the position of the neutron Fermi momentum in SNM ($k_{F,\mathrm{SNM}}=1.33 \fmiq$) and PNM ($k_{F,\mathrm{PNM}}=1.68 \fmiq$) associated with the nuclear saturation density, $\nsat^\mathrm{emp}$. 
The different curves show the results for the six Hamiltonians H1 to H7 specified in Table~\ref{tab:used_hamiltonians}.
The spread is larger in SNM (about 15~MeV) compared to PNM (about 5~MeV) because the 3N short- and intermediate-range contributions governed by $c_D$ and $c_E$ do not contribute to the PNM EOS for nonlocal regulator functions.
As expected, SNM is more attractive than PNM, as a result of the attractive contributions from the $T=0$ channels, which are absent in PNM.

\subsection{Landau mass}
\label{sec:effmass}

The momentum dependence of the nuclear interactions can be absorbed by a modification of the mass of the nucleons, which gives rise to the so-called in-medium effective mass and the Landau mass. 
Specifically, the single-particle energy can be approximated as,
\begin{equation}
\epsilon_\tau(k,n,\delta) \approx \frac{k^2}{2m_\tau^*(k,n,\delta)}+\Sigma^{(1)}(k=0,n,\delta) ,
\end{equation}
where the in-medium effective mass is defined as~\cite{Jekeunne1976},
\begin{equation}
    \frac{m^*_{\tau}(k,n,\delta)}{m_\tau} = \frac{k}{m_\tau} \bigg( \frac{d\epsilon_{\tau} (k,n,\delta)}{dk} \bigg)^{-1} \,.
    \label{eq:em}
\end{equation}
Finally, the Landau mass is defined as the effective mass~(\ref{eq:em}) taken at $k\!\!=\!\!\kf$.

\begin{figure*}
    \centering
    \includegraphics[scale=0.50]{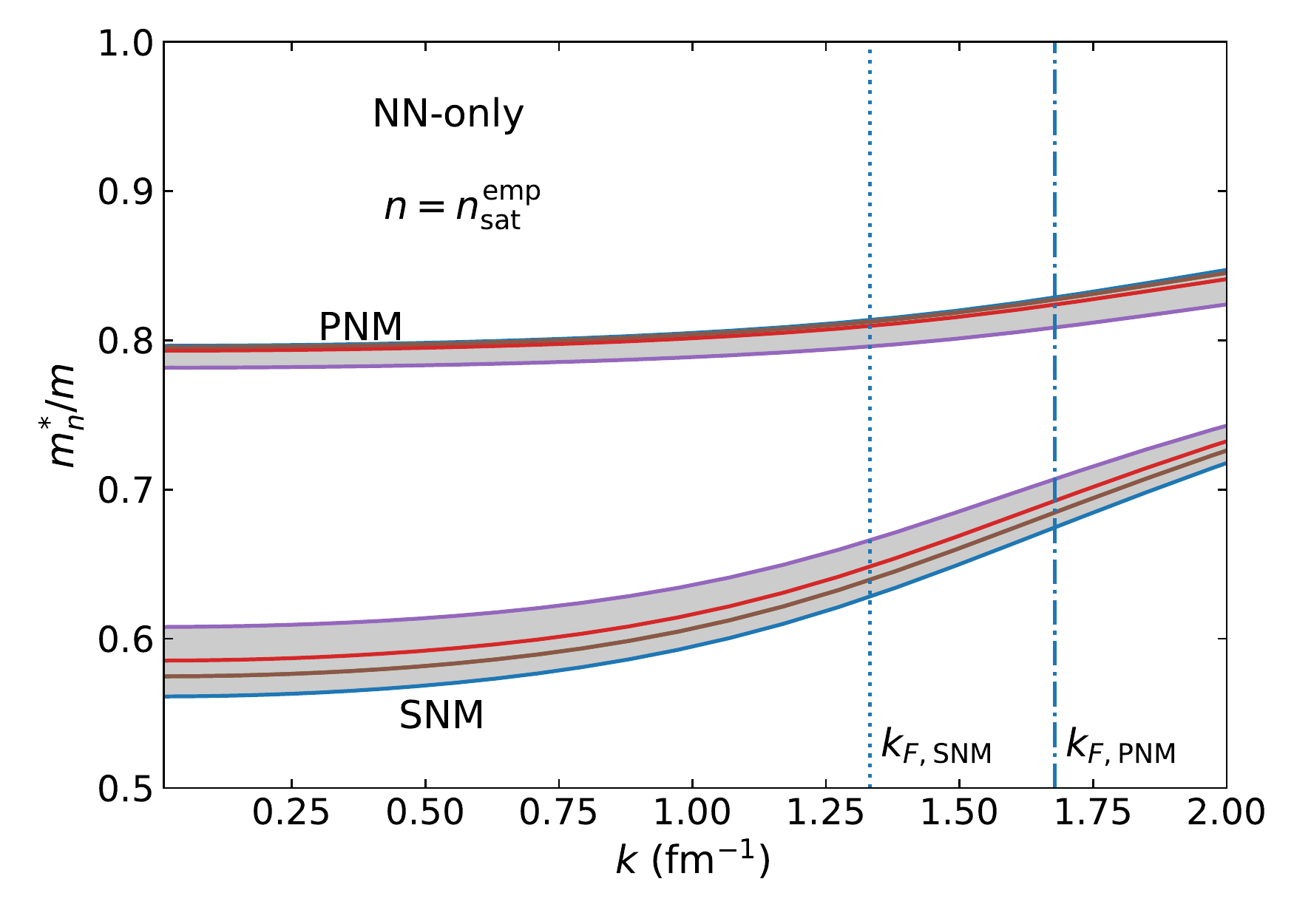}
    \includegraphics[scale=0.50]{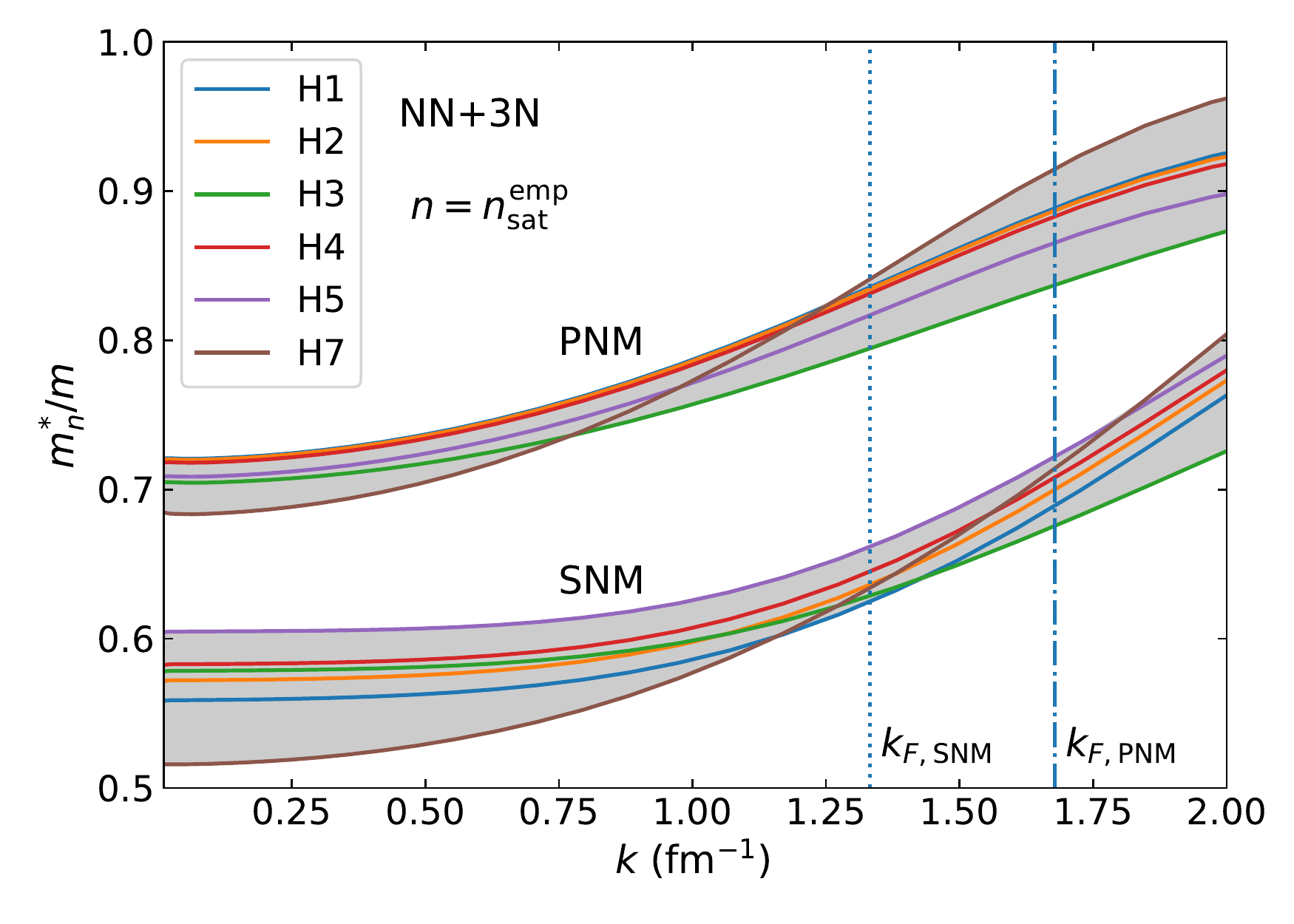}
    \caption{Same as Fig.~\ref{fig:spe} but for the neutron effective mass as function of the momentum $k$.}
    \label{fig:ms:k}
\end{figure*}

The effective masses in SNM and PNM are shown in Fig.~\ref{fig:ms:k} as functions of the momentum $k$ at a fixed density $n_{\textrm{sat}}^\mathrm{emp}$.
The effective masses are lower in SNM compared to PNM, in agreement with BHF calculations~\cite{Bombaci1991,Hassaneen2004,vanDalen2005}.
We find that the inclusion of 3N forces leads to several interesting effects on the effective mass:
(a) 3N forces generate a stronger momentum dependence compared with NN-only calculations, and (b) 3N forces have a larger impact on PNM than on SNM. 
Furthermore, the dispersion among the different Hamiltonians is slightly larger when 3N forces are included.
From Fig.~\ref{fig:ms:k}, we find for the Landau mass $m^*_n/m(\delta\!\!=\!\!0)=0.64(2)$ in SNM and $m^*_n/m(\delta\!\!=\!\!1)=0.88(4)$ in PNM when 3N forces are included.
The difference between the Landau mass in PNM and SNM at saturation density, defined as
\begin{equation}
D m^*_{n,\textrm{sat}} = m^*_n(n_{\textrm{sat}},\delta=1)-m^*_n(n_{\textrm{sat}},\delta=0) \,,
\label{eq:Landaudif}
\end{equation}
is about $D m^*_{n,\textrm{sat}} = 0.24(5)$ at $\nsat$.

\begin{figure*}
    \centering
    \includegraphics[scale=0.50]{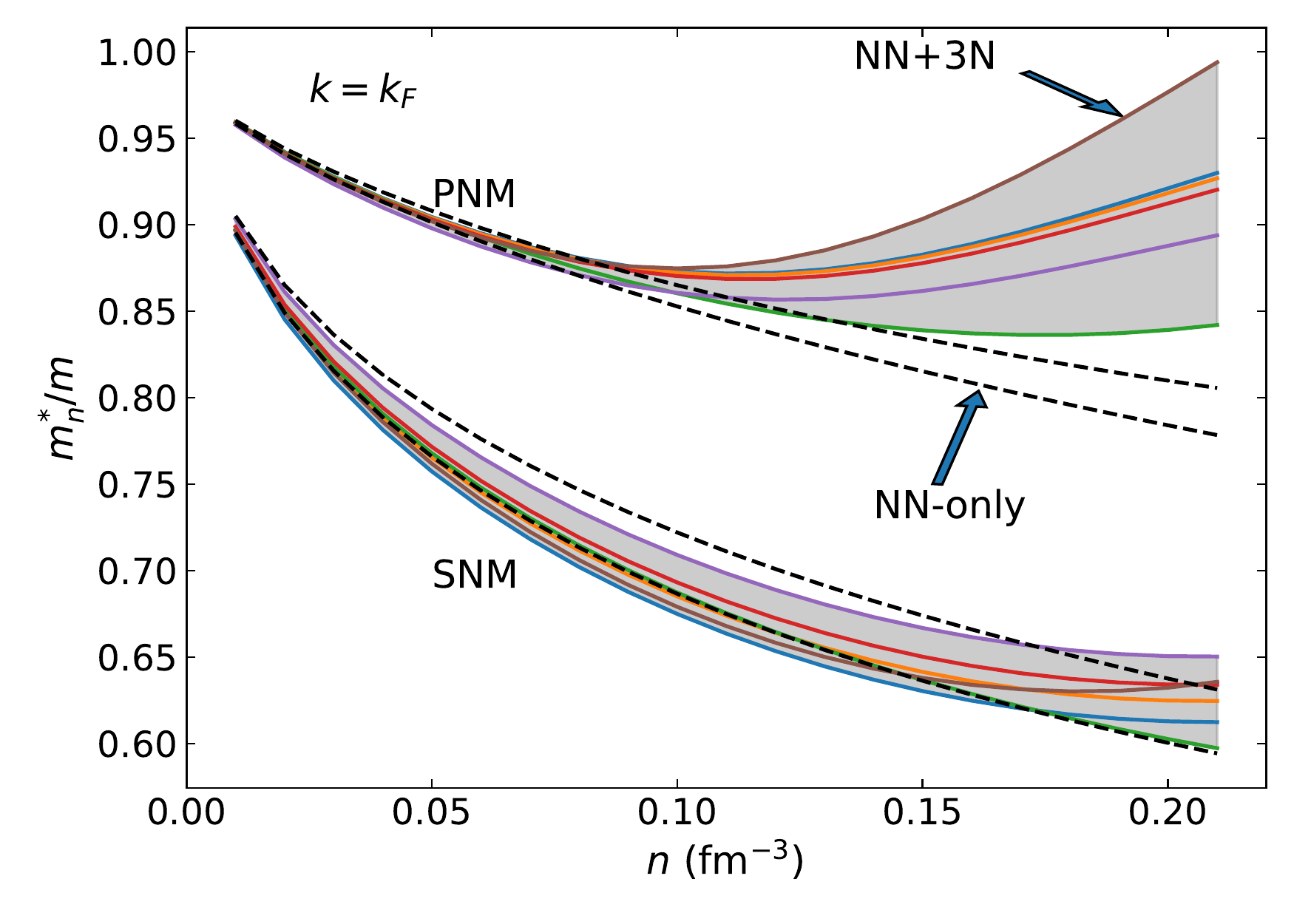}
    \includegraphics[scale=0.50]{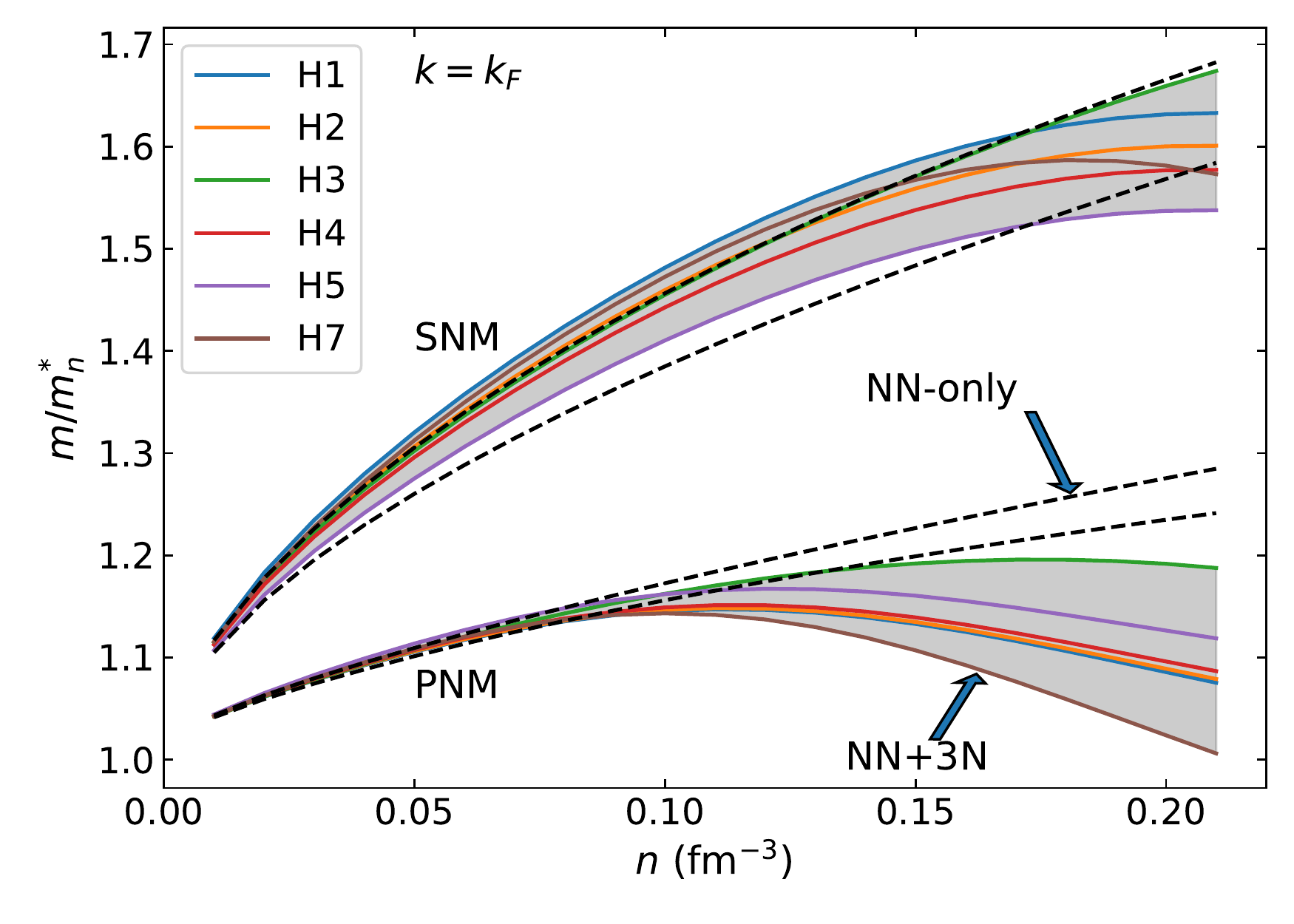}
    \caption{Landau effective mass (left) and its inverse (right) in SNM and PNM 
    as a function of the density.
    The black-dashed lines represent the upper and lower limits when only NN forces are considered, while the grey-shaded regions show the results with 3N forces included.
    The different colors correspond to the six Hamiltonians as labeled in the legend.}
    \label{fig:ms}
\end{figure*}

In Fig.~\ref{fig:ms}, we show the Landau mass (left) and its inverse (right) considering NN-only forces (dashed lines) and NN and 3N forces (gray bands) in SNM and PNM as a function of the density, $n$.
The difference of the Landau masses in PNM and SNM, $D m^*_{n}(n)$, increases with density, and is found to be about $0.24$ at saturation density, see also Fig.~\ref{fig:ms:k}.
While it is usually found that the Landau mass decreases with density~\cite{Bombaci1991, Hassaneen2004, vanDalen2005}, we find that in PNM the Landau mass first decreases at lower density, but increases again for $n>0.1\fmiq$ (except for Hamiltonian H3, which has a higher momentum cutoff applied to the 3N forces).
This effect is due to the inclusion of 3N interactions in the Hamiltonian.

Because many energy density functional (EDF) approaches approximate the inverse of the Landau mass by a linear function in density~\cite{Bender2003}, we show the inverse Landau mass in the right panel of Figure~\ref{fig:ms}.
In contrast to the EDFs approaches, we find that the density dependence of the inverse Landau mass is not linear, and that 3N forces enhance the nonlinear behavior.
To describe the inverse of the Landau mass as function of the density $n$ and asymmetry parameter $\delta$, we consider the following functional form:
\begin{eqnarray}
\left(\frac{m^{*}_{\tau}}{m}(n,\delta)\right)^{-1} &=& 1 + \left(\frac{\kappa_{\textrm{sat}}}{n_{\textrm{sat}}}+\tau_3\delta\frac{\kappa_{\textrm{sym}}}{n_{\textrm{sat}}}\right) n \nonumber \\
&&+\left(\frac{\kappa_{\textrm{sat},2}}{n_{\textrm{sat}}^2}+\tau_3\delta^2\frac{\kappa_{\textrm{sym},2}}{n_{\textrm{sat}}^2}\right) n^2\, ,\label{eq_ms}
\end{eqnarray}
where $\tau_3=1$($-1$) for neutrons (protons).
Using Eq.~(\ref{eq_ms}), we obtain in SNM and PNM, respectively,
\begin{eqnarray}
\left(\frac{m^{*}_{\textrm{SNM}}}{m}(n)\right)^{-1} &=& 1 + \frac{\kappa_{\textrm{sat}}}{n_{\textrm{sat}}} n + \frac{\kappa_{\textrm{sat},2}}{n_{\textrm{sat}}^2} n^2\, ,\label{eq_ms_SNM}\\
\left(\frac{m^{*}_{\textrm{PNM}}}{m}(n)\right)^{-1} &=& 1 + \frac{\kappa_{\textrm{PNM}}}{n_{\textrm{sat}}} n + \frac{\kappa_{\textrm{PNM},2}}{n_{\textrm{sat}}^2} n^2\, ,\label{eq_ms_PNM}
\end{eqnarray}
with $\kappa_{\textrm{PNM}}=\kappa_{\textrm{sat}}+\kappa_{\textrm{sym}}$ and $\kappa_{\textrm{PNM},2}=\kappa_{\textrm{sat},2}+\kappa_{\textrm{sym},2}$. 

The parameters $\kappa_{\textrm{sat}}$, $\kappa_{\textrm{sat},2}$, $\kappa_{\textrm{sym}}$, and $\kappa_{\textrm{sym},2}$ are obtained from fitting the expressions~\eqref{eq_ms_SNM} and \eqref{eq_ms_PNM} to the results shown in the right panel of Figure~\ref{fig:ms}.
The details of our parametric fits is discussed in Appendix~\ref{fit_procedure}. 
The relevant fit parameters, $p_{\alpha}$, are
\begin{equation}
    p_{\alpha}=\{\kappa_{\textrm{sat}}/n_{\textrm{sat}}, \kappa_{\textrm{sat},2}/n_{\textrm{sat}}^2, \kappa_{\textrm{PNM}}/n_{\textrm{sat}}, \kappa_{\textrm{PNM},2}/n_{\textrm{sat}}^2\}\,.
\end{equation}
These fit parameters are determined from the predicted Landau effective masses for each of the six Hamiltonians. 
The results of the fits for the inverse of the Landau mass are given in Table~\ref{tab:ms:fit}, where we have considered both, a linear and a quadratic fit function.
The prior distribution for each of the fit parameters is given by a normal distribution with mean $0$ and standard deviation $100$, providing an uninformative prior.
The fits are compared to three Skyrme-type interactions: NRAPR~\cite{NRAPR}, LNS5~\cite{LNS5}, and SAMI~\cite{SAMI} that satisfy the following conditions: $0.6\leqslant m^*/m(\textrm{SNM})\leqslant 0.7$, $\Delta m^*/m>0$ and $40 \MeV < L_\textrm{sym} < 60\MeV$.

\begin{table}
\centering
\tabcolsep=0.09cm
\def\arraystretch{1.9}
\caption{\label{tab:ms:fit}
Fit parameters of the inverse Landau mass considering linear and quadratic density expansions.
The fits are compared to three Skyrme-type interactions: NRAPR~\cite{NRAPR}, LNS5~\cite{LNS5}, and SAMI~\cite{SAMI}.}
\begin{tabular}{lcccc}
\hline
 & $\kappa_{\textrm{sat}}/n_{\textrm{sat}}$ & $\kappa_{\textrm{sat},2}/n_{\textrm{sat}}^2$ & $\kappa_{\textrm{PNM}}/n_{\textrm{sat}}$ & $\kappa_{\textrm{PNM},2}/n_{\textrm{sat}}^2$ \\
 & [fm$^{3}$] & [fm$^{6}$] & [fm$^{3}$] & [fm$^{6}$] \\
\hline
linear & $3.33(18)$ & - & $0.89(19)$ & - \\
quadratic & $6.25(35)$ & $-16.9(16)$ & $2.63(14)$ & $-11.1(19)$ \\
\hline
NRAPR~\cite{NRAPR} & $2.75$ & - & $1.40$ & - \\
LNS5~\cite{LNS5}  & $4.12$ & - & $2.19$ & - \\
SAMI~\cite{SAMI}  & $3.03$ & - & $2.87$ & - \\
\hline
\end{tabular}
\end{table}

\begin{figure}
\centering
\includegraphics[scale=0.53]{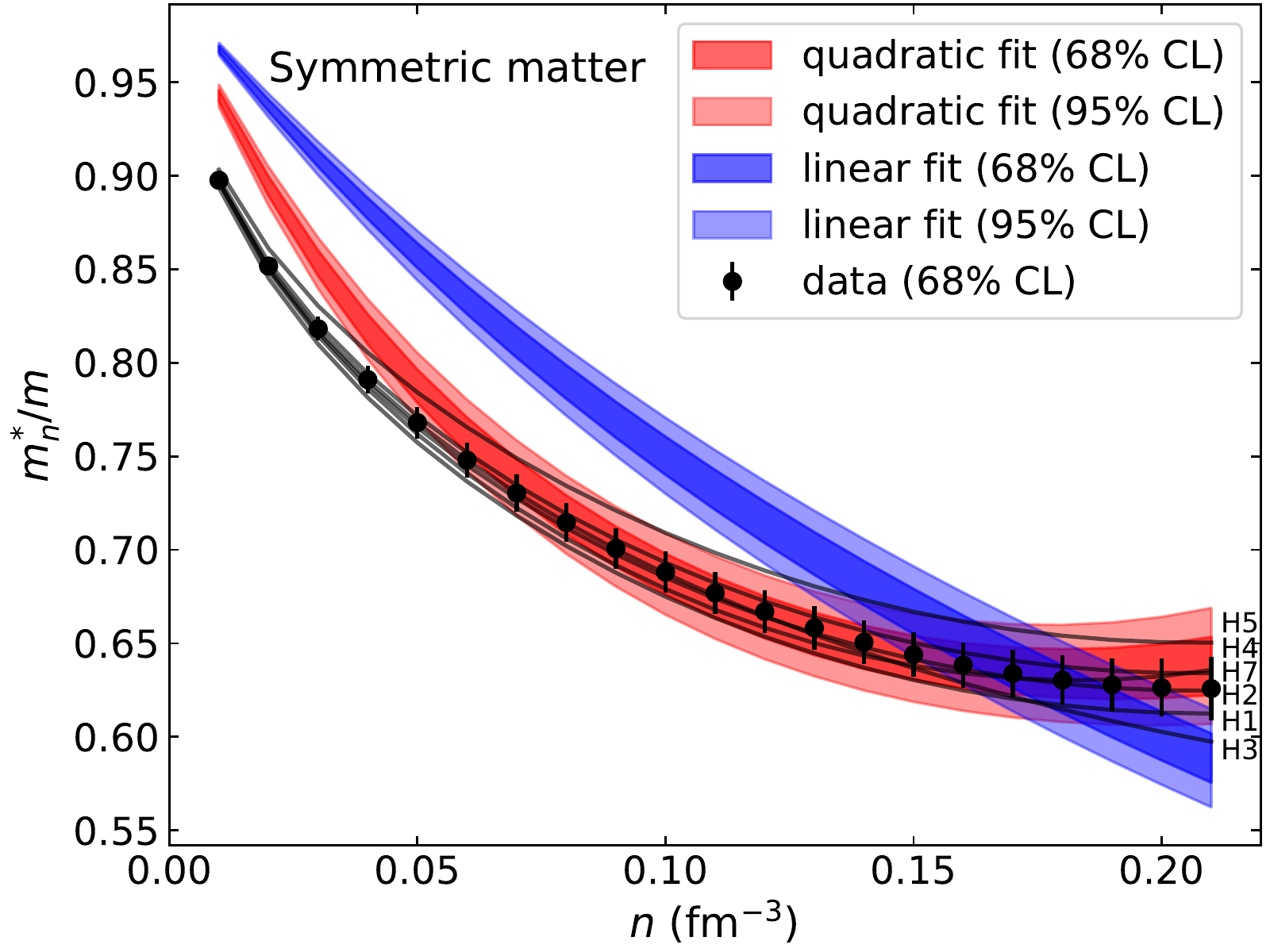}
\includegraphics[scale=0.53]{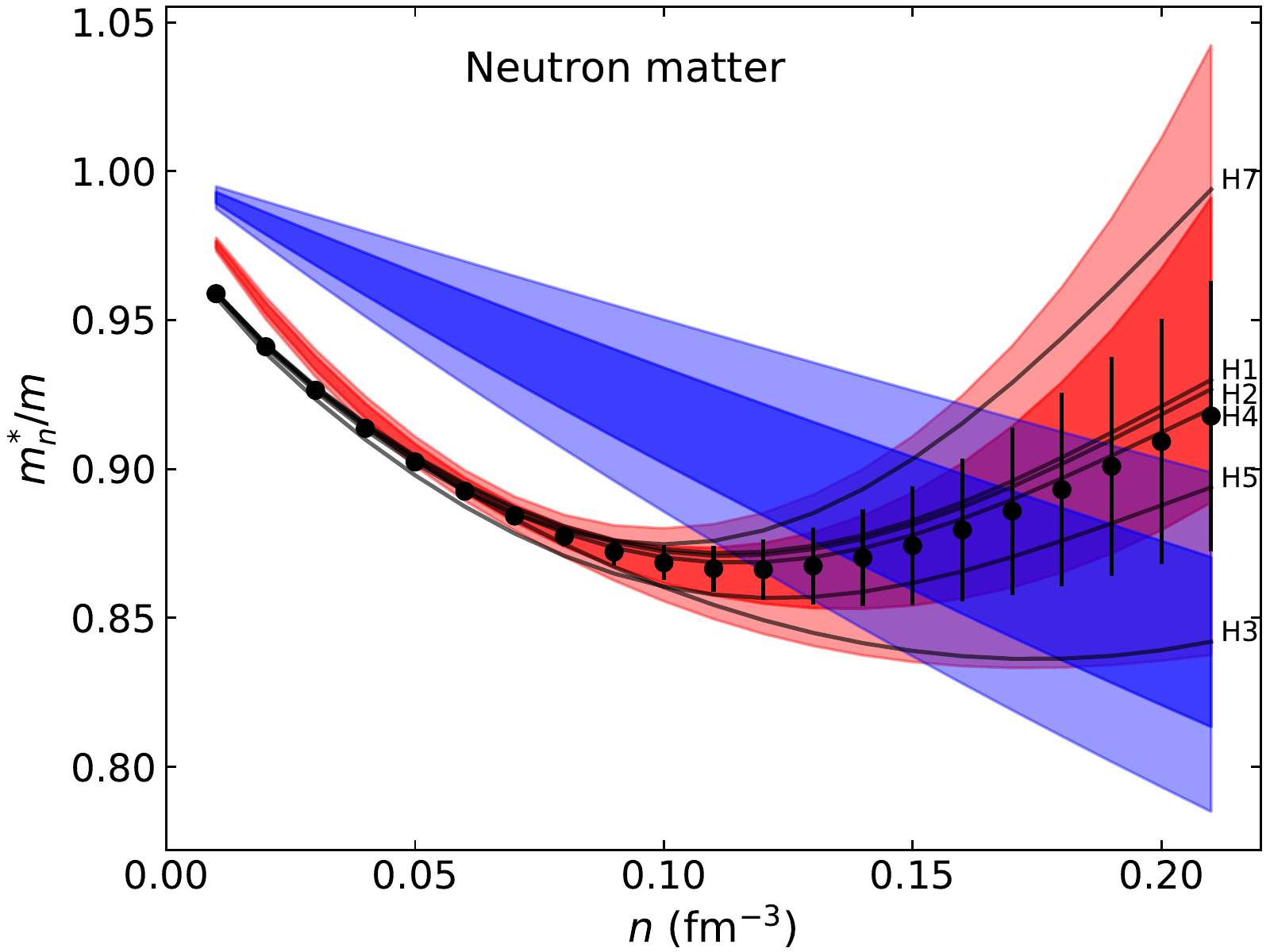}
\caption{Results of the Bayesian parametric fits of the inverse Landau mass.
The 68\% (95\%) confidence levels for the posterior distribution functions are shown as dark-shaded (light-shaded) bands.
The black lines represent the individual Hamiltonians, and the black points show the average over the six Hamiltonians with $\pm 1\sigma$ uncertainty bands.}
\label{fig:ms:fit}
\end{figure}

In Fig.~\ref{fig:ms:fit}, we compare the posterior distribution functions for the Landau mass in SNM (top panel) and PNM (bottom panel), and the input data. The predictions from the six Hamiltonians are plotted as solid lines, and, at each density, we calculate the centroid and $1\sigma$ interval given the six Hamiltonians (black points with error bars).
The $\pm1\sigma$ ($\pm2\sigma$) contours of the posterior, corresponding to the 68\% (95\%) confidence region, are depicted in red (light red) for the quadratic fit and dark blue (light blue) for the linear fit.
We fit the models to the data in the range $n=0.15-0.17\fmiq$ for the linear fit (3 data points) and from $n=0.07-0.20\fmiq$ for the quadratic fit (14 data points). 
These values are chosen to allow for the ranges to be as large as possible while, at the same time, ensuring that the fits reproduce the data around saturation density. 
While the quadratic fit performs well even outside the fit interval, down to $n \sim 0.05 \fmiq$ in SNM and PNM, the linear fit does not because of the strong curvature of the Landau mass. 
The differences between the linear and quadratic fits are further analysed in Sec.~\ref{sec:ESYM}. 

Finally, we study the splitting of the neutron and proton Landau masses in ANM, defined as
\begin{equation}
\Delta m^*_{\textrm{sat}}(\delta) = m^*_n(n_{\textrm{sat}},\delta)-m^*_p(n_{\textrm{sat}},\delta) \, .
\end{equation}
In PNM ($\delta=1$), this splitting can be expressed in terms of the difference $D m^*_{\textrm{sat}}$, see Eq.~\eqref{eq:Landaudif}, as
\begin{equation}
\frac{\Delta m^*_\textrm{sat}}{m}(\textrm{PNM}) \approx
\frac{D m^*_{n,\textrm{sat}}}{m} + \mathcal{O}\left(\left(\frac{\kappa_{\textrm{sym}}+\kappa_{\textrm{sym},2}}{1+\kappa_{\textrm{sat}}+\kappa_{sat,2}}\right)^2\right) \, .
\end{equation}
From our fits, we can estimate that the neglected terms account for about 5\% of the splitting (more precisely, 7\% for the linear fit of the effective mass, and 3\% for the quadratic fit), which is small considering the present uncertainty of this quantity.
The splitting of the Landau mass is, thus, approximately given by the difference of the Landau mass between PNM and SNM.
The splitting of the Landau mass obtained here is compatible with the one obtained in the literature for BHF~\cite{Bombaci1991,Zuo2001,Zuo2002} and Dirac-BHF~\cite{Hassaneen2004,vanDalen2005} approaches.

\section{Energy expansion in the isospin asymmetry parameter $\delta$}
\label{sec:expansion}

A general expression for the expansion~\eqref{eq:EA_quadratic_expansion}
of energy observables in nuclear matter was suggested in Eq.~(27) of Ref.~\cite{Wellenhofer2016}. 
From this expression, we consider all contributions up to $\delta^4$, including the logarithmic term, and rewrite it as
\begin{align}
y(n,\delta)&\approx y_{\textrm{SNM}}(n)+y_{\textrm{sym},2}(n)\delta^2+y_{\textrm{sym},4}(n)\delta^4 \nonumber \\
&\quad+y_{\textrm{sym,log}}(n)\delta^4\log\vert\delta\vert \, .
\label{eq:general}
\end{align}
The term $y_{\textrm{sym},\textrm{log},4}\delta^4\log\vert\delta\vert$ originates from the second-order contribution in the many-body expansion in the neutron-proton channel, as explained in Ref~\cite{Wellenhofer2016}.
The corresponding contribution to the symmetry energy $y_{\textrm{sym}}$ is defined as
\begin{eqnarray}
y_{\textrm{sym}}(n) &=& y_{\textrm{PNM}}(n)-y_{\textrm{SNM}}(n) \,,
\label{eq:general:sym}
\end{eqnarray}
where $y_{\textrm{SNM}}(n)=y(n,\delta=0)$ and $y_{\textrm{PNM}}(n)=y(n,\delta=1)$.
The non-quadratic contribution to the symmetry energy is defined as
\begin{eqnarray}
y_{\textrm{sym,nq}}(n) &=& y_{\textrm{sym}}(n) - y_{\textrm{sym,2}}(n)  
\,.
\label{eq:general:symnq}
\end{eqnarray}
Note, that since the logarithmic term vanishes in SNM and PNM, it also does not contribute to the non-quadratic term~\eqref{eq:general:symnq}.

The quantity $y$ in Eq.~\eqref{eq:general} can be the energy per particle $e$, as originally suggested by Kaiser~\cite{Kaiser2015}, or any other energy contribution can be expanded as well. 
For instance, as discussed in Sec.~\ref{sec:introduction}, one can expand the kinetic energy $t$, which coincides with the FFG energy given in Eq.~\eqref{eq:FFG}, $t=e^\text{FFG}$.
The $\delta$-expansion of the kinetic energy up to fourth order is given by
\begin{equation}
t(n,\delta) \approx t_{\textrm{SNM}}(n) + t_{\textrm{sym},2}(n)\delta^2 + t_{\textrm{sym},4}(n)\delta^4\, ,
\end{equation}
with
\begin{align}
t_{\textrm{SNM}}(n) &= t_{\textrm{SNM}}^{\textrm{sat}}\left(\frac{n}{n_{\textrm{sat}}}\right)^{2/3}\, ,\\
t_{\textrm{sym},2}(n) &= \frac{5}{9}t_{\textrm{SNM}}^{\textrm{sat}}\left(\frac{n}{n_{\textrm{sat}}}\right)^{2/3}\, , \\
t_{\textrm{sym},4}(n) &= \frac{5}{243}t_{\textrm{SNM}}^{\textrm{sat}}\left(\frac{n}{n_{\textrm{sat}}}\right)^{2/3}\, .
\end{align}
We note that $t_{\textrm{sym, log}}=0$ for the kinetic energy because it is a one-body operator.
At saturation density, one finds $t_{\textrm{sym},4}(n_{\textrm{sat}})\approx 0.45$~MeV, while the contribution of all higher-order terms starting with $t_{\textrm{sym,6}}$ sum up to about 0.25~MeV.

When including nuclear interactions, the total energy is given by the sum of the kinetic energy $t(n,\delta)$ and the potential energy $e^{\textrm{pot}}(n,\delta)$.
The potential energy can, thus, be defined as
\begin{equation}
e^{\textrm{pot}}(n,\delta) = e(n,\delta) - t(n,\delta) \,,
\end{equation}
where the total energy $e(n,\delta)$ is obtained from the MBPT calculations~\cite{Drischler:2015eba} and the kinetic energy is analytic.
Again, the potential energy can be expanded using Eq.~\eqref{eq:general}, 
\begin{align}
e^{\textrm{pot}}(n,\delta)&\approx e^{\textrm{pot}}_{\textrm{SNM}}(n)+e^{\textrm{pot}}_{\textrm{sym},2}(n)\delta^2
+e^{\textrm{pot}}_{\textrm{sym},4}(n)\delta^4 \nonumber \\
&\quad +e^{\textrm{pot}}_{\textrm{\textrm{sym},log},4}(n)\delta^4\log\vert\delta\vert \,.
\end{align}

Another way to express the total energy is to explicitly include the in-medium correction to the nucleon mass, modifying the kinetic and potential energies but not their sum,
\begin{equation}
e(n,\delta)=t^*(n,\delta)+e^{\textrm{pot}*}(n,\delta)\,,
\end{equation}
where $t^*(n,\delta)$ is the effective kinetic energy calculated with the Landau mass $m^*_\tau(\delta)$.
Here, $\tau=n$ or $p$ for neutrons and protons, respectively, and $e^{\textrm{pot}*}(n,\delta)$ is the effective potential energy.
We refer to Sec.~\ref{sec:effmass} for more details on the Landau mass.
The effective kinetic energy is defined as
\begin{eqnarray}
t^{*}(n,\delta) &=& \frac{t_{\textrm{SNM}}^{\textrm{sat}}}{2}\left(\frac{n}{n_{\textrm{sat}}}\right)^{2/3} \Big[ \frac{m}{m^*_n(\delta)}(1+\delta)^{5/3}\nonumber \\
&&\hspace{2cm}+\frac{m}{m^*_p(\delta)}(1-\delta)^{5/3} \Big] \, ,
\label{eq:tkinstar}
\end{eqnarray}
and the effective potential energy is given by
\begin{align}
e^{\textrm{pot}*}(n,\delta) &= e(n,\delta) - t^{*}(n,\delta) \\\nonumber&= t(n,\delta) - t^{*}(n,\delta) + e^{\textrm{pot}}(n,\delta) \, .
\label{eq:epoteff}
\end{align}
Similarly to $e$ and $e^{\textrm{pot}}$, the effective potential energy $e^{\textrm{pot}*}$ can be expanded using Eq.~\eqref{eq:general},
\begin{align}
e^{\textrm{pot}*}(n,\delta)&\approx e^{\textrm{pot}*}_{\textrm{SNM}}(n)+e^{\textrm{pot}*}_{\textrm{sym},2}(n)\delta^2+e^{\textrm{pot}*}_{\textrm{sym},4}(n)\delta^4 \nonumber \\
&\quad+e^{\textrm{pot}*}_{\textrm{\textrm{sym},log}}(n)\delta^4\log\delta \, .
\end{align}

In the following, we use these notations for analyzing the $\delta$-dependence of the total, potential, and effective potential energies.

\section{Meta-Model for Symmetric and Neutron matter}
\label{sec:SNMPNM}

To describe the MBPT data for the energy per particle in SNM and PNM, we use in this work a functional form described by a meta-model (MM) for nuclear matter similar to the one suggested in Ref.~\cite{Margueron2018a}, but generalized to a potential energy with non-quadratic $\delta$ dependence. 
The MM is adjusted to MBPT data sampled on a given grid in the asymmetry parameter $\delta$~\cite{Drischler:2015eba}.
This is in contrast to Wellenhofer~\etal~\cite{Wellenhofer2016}, who used a finite difference method~\cite{Fornberg1988} on an adjustable grid to determine all derivatives with respect to $\delta$ of interest. 
The MM, instead, provides a flexible polynomial-type approach to nuclear matter, which allows us to accurately determine the higher-order coefficients in the $\delta$ expansion, even for the fixed grid considered here.

For SNM and PNM, the energy per particle in the MM reads
\begin{eqnarray}
e_{\textrm{SNM}}^\mathrm{MM}(n) &=& t_{\textrm{SNM}}^*(n) + e^{\textrm{pot}*}_{\textrm{SNM}}(n) \, , \label{eq_fit_SNM}\\
e_{\textrm{PNM}}^\mathrm{MM}(n) &=& t_{\textrm{PNM}}^*(n) + e^{\textrm{pot}*}_{\textrm{PNM}}(n) \,.
\end{eqnarray}
The kinetic energy is determined from Eq.~(\ref{eq:tkinstar}) with the Landau mass, see Sec.~\ref{sec:effmass}.
The potential energies are expanded about $\nsat$ in terms of the parameter
$$
x \equiv \frac{n-n_{\textrm{sat}}}{3n_{\textrm{sat}}}
$$
as follows
\begin{eqnarray}
e^{\textrm{pot}*}_{\textrm{SNM}}(n) &=& \sum_{j=0}^{N} \frac{1}{j!}v_{\textrm{SNM},j} x^j + v_{\textrm{SNM}}^\mathrm{low-n} \,x^{N+1} e^{-b_{\textrm{sat}}\frac{n}{n_\mathrm{sat}^\mathrm{emp}}} \label{eq:vsat}\, , \nonumber \\
e^{\textrm{pot}*}_{\textrm{PNM}}(n) &=& \sum_{j=0}^{N} \frac{1}{j!}v_{\textrm{PNM},j} x^j + v_{\textrm{PNM}}^\mathrm{low-n} \,x^{N+1} e^{-b_{\textrm{PNM}}\frac{n}{n_\mathrm{sat}^\mathrm{emp}}}\, ,\nonumber  \label{eq:vPNM}
\end{eqnarray}
where $b_{\textrm{PNM}} =b_{\textrm{sat}}+b_{\textrm{sym}}$, and the second term on the right is a low-density correction. 
This correction represents the low-density contribution of all higher-order terms neglected in the summation, and scales like $x^{N+1}$ at leading order, where $N$ is the upper limit of the power in the density expansion.
In the original nucleonic MM of Ref.~\cite{Margueron2018a}, the low-density EOS correction was simply parameterized by a fixed coefficient $b=b_{\textrm{sat}}=b_\textrm{PNM}\approx 6.93$.
In the improved MM considered here, we introduce two parameters ($b_\mathrm{sat}$ and $b_\mathrm{sym}$) controlling the density dependence of the low-density corrections in PNM and SNM separately.
It was suggested that considering an expansion up to $N=4$ allows the reproduction of the pressure and sound speed of about 50 known EDF up to about $4n_\mathrm{sat}$, see Ref.~\cite{Margueron2018a} for more details. 
In principle, it is not necessary to consider such a high $N$ in the present analysis. 
The inclusion of high-order contributions, however, affects the determination of the low-order ones, as discussed in Ref.~\cite{Margueron2019}, even if the data does not constrain the high-order terms themselves.

Imposing that the energies per particle vanish at \mbox{$n=0\fmiq$}, we obtain the following relations
\begin{eqnarray}
e^{\textrm{pot}*}_{\textrm{SNM}}(n) &=& \sum_{j=0}^{N} \frac{1}{j!} v_{\textrm{SNM},j} \, x^j \, u_{\textrm{SNM},j}(x) \, ,\\
e^{\textrm{pot}*}_{\textrm{PNM}}(n) &=& \sum_{j=0}^{N} \frac{1}{j!} v_{\textrm{PNM},j} \, x^j \, u_{\textrm{PNM},j}(x)\, ,
\end{eqnarray}
where
\begin{eqnarray}
u_{\alpha,j}(x)=1-(-3x)^{N+1-j}e^{-b_{\alpha} n / n_{\textrm{sat}}^\mathrm{emp}}\,,
\end{eqnarray}
and $\alpha$ indicates either SNM or PNM and the corresponding $b_{\textrm{sat}}$ or $b_{\textrm{PNM}}$.

In the MM, the coefficients $v_{\alpha,1}$ to $v_{\alpha,N}$ are related to the nuclear empirical parameters (NEPs), such as $E_{\textrm{sat}}$, $K_{\textrm{sat}}$, $E_{\textrm{sym}}$, $L_{\textrm{sym}}$, etc. The NEPs for SNM are defined by the density expansion
\begin{eqnarray}
e_{\textrm{SNM}}(n) &= E_{\textrm{sat}} + \frac 1 2 K_{\textrm{sat}} x^2 + \frac 1 6 Q_{\textrm{sat}} x^3 \, \nonumber \\ &+ \frac 1 {24} Z_{\textrm{sat}} x^4 + \dots\,,
\end{eqnarray}
whereas the NEPs for PNM are defined by
\begin{eqnarray}
e_{\textrm{PNM}}(n) &= E_{\textrm{PNM}} + L_{\textrm{PNM}} x + \frac 1 2 K_{\textrm{PNM}} x^2 \, \nonumber \\ &+ \frac 1 6 Q_{\textrm{PNM}} x^3 + \frac 1 {24} Z_{\textrm{PNM}} x^4 + \dots\;.
\end{eqnarray}
We use the following relations between the MM parameters and the NEPs for the isoscalar parameters controlling the SNM EOS:
\begin{eqnarray}
v_{\textrm{SNM},0} &=& E_{\textrm{sat}} - t_{\textrm{SNM}} (1 + \kappa_{\textrm{sat}} + \kappa_{\textrm{sat},2})\, ,  \\
v_{\textrm{SNM},1} &=& -t_{\textrm{SNM}} (2 + 5\kappa_{\textrm{sat}} +
8\kappa_{\textrm{sat},2})\,,  \nonumber \\
v_{\textrm{SNM},2} &=& K_{\textrm{sat}} - 2 t_{\textrm{SNM}} (-1 + 5\kappa_{\textrm{sat}} +
20\kappa_{\textrm{sat},2})\,,  \nonumber \\
v_{\textrm{SNM},3} &=& Q_{\textrm{sat}} - 2 t_{\textrm{SNM}} (4 -5\kappa_{\textrm{sat}} +
40\kappa_{\textrm{sat},2})\,,  \nonumber \\
v_{\textrm{SNM},4} &=& Z_{\textrm{sat}} -8 t_{\textrm{SNM}} (-7 + 5\kappa_{\textrm{sat}} - 10\kappa_{\textrm{sat},2})\,, \nonumber
\end{eqnarray}
while the isovector parameters describing the PNM EOS are
\begin{eqnarray}
v_{\textrm{PNM},0} &=& E_{\textrm{PNM}} - 2^{\frac{2}{3}} t_{\textrm{SNM}} (1 + \kappa_{\textrm{PNM}} + \kappa_{\textrm{PNM},2})\,,  \\
v_{\textrm{PNM},1} &=& L_{\textrm{PNM}}- 2^{\frac{2}{3}}  t_{\textrm{SNM}} (2 + 5\kappa_{\textrm{PNM}} +
8\kappa_{\textrm{PNM},2})\,, \nonumber  \\
v_{\textrm{PNM},2} &=& K_{\textrm{PNM}} - 2^{\frac{5}{3}} t_{\textrm{SNM}} (-1 + 5\kappa_{\textrm{PNM}} + 20\kappa_{\textrm{PNM},2})\,, \nonumber  \\
v_{\textrm{PNM},3} &=& Q_{\textrm{PNM}} - 2^{\frac{5}{3}} t_{\textrm{SNM}} (4 -5\kappa_{\textrm{PNM}} +  40\kappa_{\textrm{PNM},2})\,, \nonumber  \\
v_{\textrm{PNM},4} &=& Z_{\textrm{PNM}} - 2^{\frac{11}{3}} t_{\textrm{SNM}} (-7 + 5\kappa_{\textrm{PNM}} - 10\kappa_{\textrm{PNM},2})\,.\nonumber
\end{eqnarray}

These relations represent another difference to the original nucleonic MM of Ref.~\cite{Margueron2018a}, where the isovector coefficients were determined assuming a quadratic isospin-asymmetry dependence of the symmetry energy.
The isovector contribution of the present MM is built on the global symmetry energy~\eqref{eq:esym1}, which allows for possible non-quadratic contributions to the symmetry energy.
These contributions will be estimated from the difference between the global symmetry energy and its quadratic contribution, as detailed in Sec.~\ref{sec:ESYM}.

In the MM used here, there are five NEP in SNM, including $n_\mathrm{sat}$, and five additional NEP in PNM.
Considering the two parameters controlling the low-density EOS, $b_{\textrm{sat}}$ and $b_{\textrm{sym}}$, there is a total of 12 parameters that need to be determined.
Note, that these parameters carry uncertainties that reflect the current lack of knowledge of the nuclear EOS.
In our Bayesian fits, we use the priors for the NEP from the analysis presented in Refs.~\cite{Margueron2018a,Margueron2018b} and summarized in Table~\ref{tab:tbl_2}.
Here, we additionally vary the parameters $b_{\textrm{sat}}$ and $b_\textrm{PNM}$ to reproduce the low-density behavior of the energy per particle in SNM and PNM.

\begin{table*}
\centering
\tabcolsep=0.40cm
\def\arraystretch{1.9}
\caption{Priors and posteriors of the NEP from analyses of SNM and PNM.
We report results for the different scalings described in the text.
Values within parentheses represent the error bars at the $\pm 1\sigma$ level.
NEPs for the following three Skyrme-type interactions are given: NRAPR~\cite{NRAPR}, LNS5~\cite{LNS5} and SAMI~\cite{SAMI}.}
\label{tab:tbl_2}
\begin{tabular}{cccccccc}
\hline
 Scaling & $n_{\textrm{sat}}$  & $E_{\textrm{sat}}$ &  & $K_{\textrm{sat}}$ & $Q_{\textrm{sat}}$ & $Z_{\textrm{sat}}$ & $b_{\textrm{sat}}$ \\
 & (fm$^{-3}$) & (MeV) & & (MeV) & (MeV) & (MeV) & \\
\hline
 prior & $0.160(10)$ & $-15.50(100)$ & & $230(20)$ & $-300(400)$ & $1300(500)$ & $0(50)$ \\
1  & $0.166(8)$ & $-15.48(58)$ & & $211(14)$ & $-573(133)$  & $1055(474)$ & $17(5)$ \\
 2  & $0.163(8)$ & $-15.07(57)$ & & $227(18)$ & $-172(243)$ & $1287(499)$ & $9(5)$ \\
 3  & $0.163(8)$ & $-15.07(57)$ & & $227(18)$ & $-172(243)$ & $1287(499)$ & $9(5)$ \\
 3* & $0.161(7)$   & $-15.17(57)$ & & $226(18)$ & $-306(186)$  & $1324(497)$ & $17$$\dagger$  \\
 \hline
 NRAPR~\cite{NRAPR} & $0.161$ & $-15.85$ &  & $226$ & $-363$ & $1611$ \\
 LNS5~\cite{LNS5}   & $0.160$ & $-15.57$ &  & $240$ & $-316$ & $1255$ \\
 SAMI~\cite{SAMI}   & $0.159$ & $-15.93$ &  & $245$ & $-339$ & $1330$ \\
\hline
 Scaling & $n_{\textrm{sat}}$ & $E_{\textrm{PNM}}$ & $L_{\textrm{PNM}}$ & $K_{\textrm{PNM}}$ & $Q_{\textrm{PNM}}$ & $Z_{\textrm{PNM}}$ & $b_{\textrm{PNM}}$ \\
 & (fm$^{-3}$) & (MeV) & (MeV) & (MeV) & (MeV) & (MeV) &  \\
\hline
 prior & - & $16.00(300)$ & $50(10)$ & $100(100)$ & $0(400)$ & $-500(500)$ & $0(50)$ \\
 1  & $0.166(8)$$\dagger\dagger$ & $16.61(93)$ & $48(5)$ & $40(37)$ & $-320(224)$ & $-388(494)$ & $42(4)$ \\
 2  & $0.163(8)$$\dagger\dagger$  & $16.30(93)$ & $47(5)$  & $75(40)$ & $34(285)$ & $-504(497)$ & $15(9)$ \\
 3  & $0.163(8)$$\dagger\dagger$  & $16.30(93)$ & $47(5)$  & $75(40)$ & $34(285)$ & $-504(497)$ & $15(9)$ \\
 3* & $0.161(7)$$\dagger\dagger$  & $16.16(89)$ & $46(5)$  & $57(34)$ & $-110(206)$ & $-450(492)$  & $42$$\dagger$  \\
 \hline
NRAPR~\cite{NRAPR} & $0.161$ & $18.33$ & $65$ & $108$ & $-52$ & $-236$ \\
LNS5~\cite{LNS5}   & $0.160$ & $15.29$ & $57$ & $130$ & $-34$ & $-416$ \\
SAMI~\cite{SAMI}   & $0.159$ & $13.32$ & $47$ & $127$ &  $35$ & $-873$ \\
\hline
\end{tabular}

$\dagger$ Fixed parameter. $\dagger\dagger$ Quoted values are the $n_{\textrm{sat}}$ priors considered in PNM and obtained from SNM posteriors.
\end{table*}

\begin{figure*}
    \centering
    \includegraphics[scale=0.47]{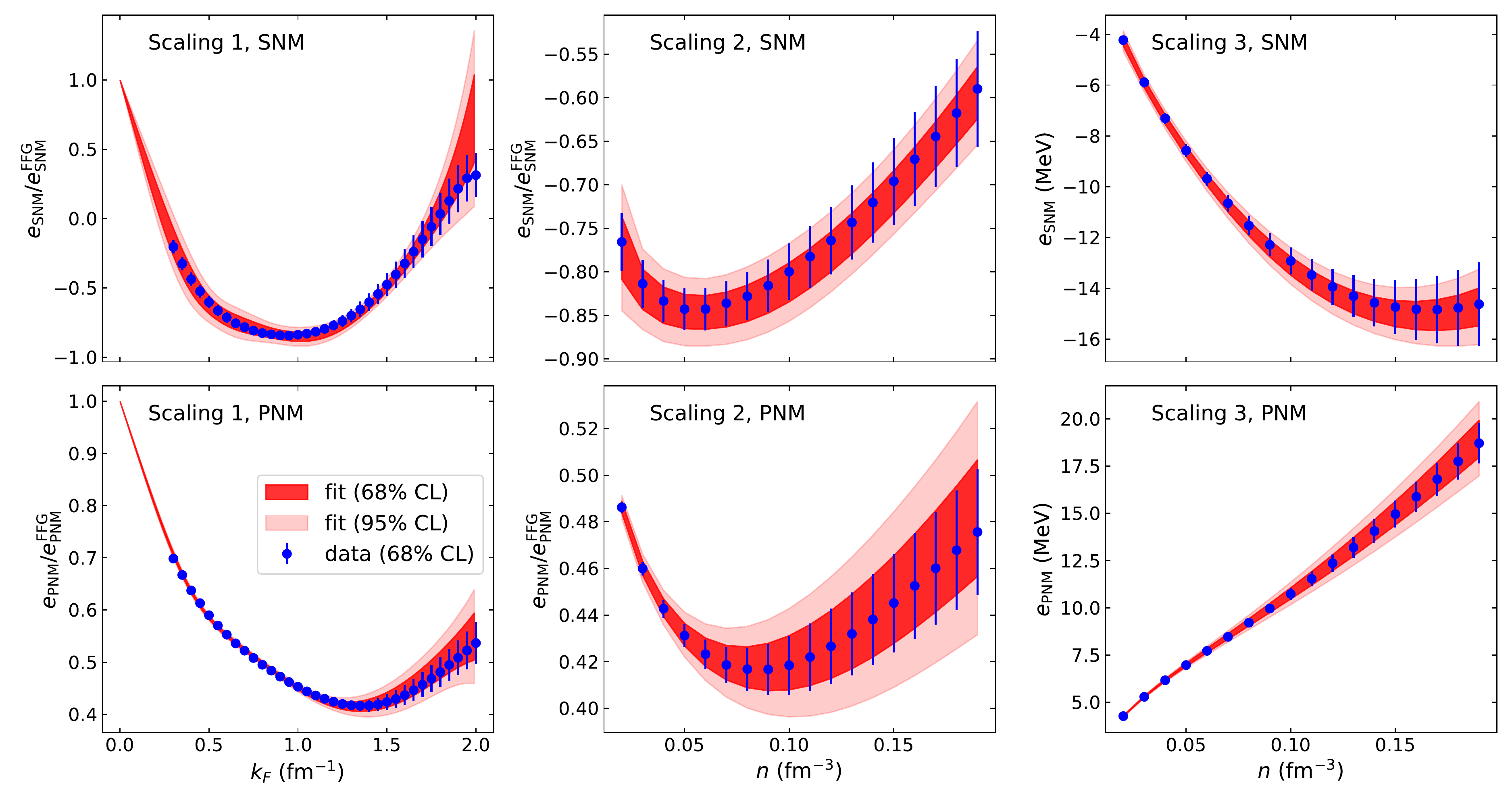}
    \caption{Comparison of the Bayesian inference results for the MM of this work (red bands) with the MBPT data (blue points) for SNM (top panels) and PNM (bottom panels) and the three different scalings described in the main text. 
    The bands are given at 65\% (dark-shaded) and 95\% confidence level (light-shaded), whereas the data points are shown with the $\pm 1\sigma$ uncertainty estimate.}
    \label{fig:scalings}
\end{figure*}

We show the fitted parameters in Table~\ref{tab:tbl_2}.
Note, that both the posteriors and priors given in the table represent the means of normal distributions with the standard deviations given in parenthesis.
The posteriors we obtain for the NEPs may depend on the exact representation of the data points, i.e., if the data is equidistant in $n$ or $\kf$.
To gauge the sensitivity to this choice, we investigate in the following three possible data representations, here called scalings.
We show the results for all scalings in Fig.~\ref{fig:scalings}.

Since we require our fit to provide a faithful representation of low-density nuclear matter (with a fair weight for the low-density points) in order to fix $b_{\textrm{sat}}$ and $b_{\textrm{PNM}}$, we adopt for scaling~1 the representation of $e/e^\textrm{FFG}$ as a function of the Fermi momentum $\kf$.
Scaling~1 provides the best representation to analyze the low-density properties of the energy per particle because
an equidistant grid in $\kf$ leads to a very dense data set at low densities.
Note that, as mentioned in Sec.~\ref{sec:Micro}, the original MBPT data~\cite{Drischler:2015eba} is provided on an equidistant grid in $\kf$.
The scaling of the $y$-axis normalizes the energies to the same order of magnitude at all $\kf$.
For scaling~2, we choose the representation of $e/e^\mathrm{FFG}$ on an equidistant grid in density.
We use a cubic spline to interpolate the energies per particle from the original Fermi momentum grid to a equally spaced density grid.
By switching the equidistant grid in momenta to densities, scaling~2 reduces the weight for the low-density data points and, therefore, is more appropriate to fit the NEPs, which are determined around saturation density.
Finally, scaling~3 represents the energy per particle on an equidistant grid in density, as it is more often presented in the literature. 
Hence, the only difference to scaling~2 is the scaling of the $y$-axis.
The results for each of the three scalings are shown in Fig.~\ref{fig:scalings} for SNM and PNM, while the posteriors for the NEPs are given in Table~\ref{tab:tbl_2}.
Note, that the NEP $n_{\textrm{sat}}$ is only meaningful in SNM, while its uncertainty influences the determination of the NEPs in PNM.
In our approach, we therefore vary $n_{\textrm{sat}}$ in PNM within the posterior uncertainty obtained from the fit in SNM.
In this way, the NEP in PNM naturally include the uncertainty in $n_{\textrm{sat}}$.

Scaling~1 is ideal for analyzing the low-density behavior of the energy per particle.
When simultaneously varying the 12 MM parameters, we find $b_{\textrm{SNM}}=17(5)$ and $b_{\textrm{PNM}}=42(4)$ as well as the values for the 10 NEP given in Table~\ref{tab:tbl_2}.
The density dependence of the low-density correction is, thus, very different in SNM and PNM, in contrast to the original MM of Ref.~\cite{Margueron2018a}.
Illustrating the importance of the scaling choice, we find some differences between the NEPs obtained from scaling~1 and scalings~2 and~3.
These differences are usually small compared to the uncertainties, except for $Q_{\textrm{sat}}$ in SNM, as well as $Q_{\textrm{PNM}}$ in PNM.
We note that the fits from scalings~2 and~3 are identical and, hence, the scaling of the energy with respect to $e^\mathrm{FFG}$ has a negligible effect.

Finally, we fix the values for $b_{\textrm{SNM}}$ and $b_{\textrm{PNM}}$ from scaling~1, and re-fit all remaining NEP considering scaling~3.
The results are referred to as scaling~$3^*$.
Fixing $b_{\textrm{SNM}}$ and $b_{\textrm{PNM}}$ has the largest impact on $Q_{\textrm{sat}}$ and $Q_{\textrm{PNM}}$, as expected, but differences between scaling~$3$ and~$3^*$ are small compared to the overall uncertainties.
We, therefore, conclude that the parameters $b_{\textrm{SNM}}$ and $b_{\textrm{PNM}}$ do not have a significant impact on the determination of the NEPs, and can be fixed from the fit to low-density matter (scaling~1).
We stress that the higher-order NEPs $Z_{\textrm{sat}}$ and $Z_{\textrm{PNM}}$ are not constrained by our data, because the density range of the MBPT data is limited to densities below $0.2~\fmiq$.
However, they contribute to the uncertainty of the other NEPs~\cite{Margueron2019}.

For the NEPs describing nuclear saturation we obtain $\nsat=0.161(7)\fmiq$, $E_\mathrm{sat}=-15.17(57)$~MeV, and $K_\mathrm{sat}=226(18)$~MeV. 
The results are
consistent with the original analysis in Ref.~\cite{Drischler:2015eba}, which obtained $\nsat=0.143-0.190\fmiq$, $E_\mathrm{sat}=-(15.1-18.3)$~MeV, and $K_\mathrm{sat}=223-254$~MeV using a Hartree-Fock spectrum. 
However, our uncertainties are generally smaller because we explicitly guide our fits in Fig.~\ref{fig:scalings} by empirical (or ``expert'') knowledge~\cite{Margueron2018a} through prior distributions of the fit parameters. 
In PNM, where empirical constraints are lacking, the fits are therefore closer to the MBPT data.



\section{Symmetry energy}
\label{sec:ESYM}

Using the results obtained in Secs.~\ref{sec:effmass} and \ref{sec:SNMPNM}, we now determine the properties of the symmetry energy and the relative contributions of the quadratic and non-quadratic terms.

\subsection{Global symmetry energy $e_{\textrm{sym}}$}

\begin{figure*}
\centering
\includegraphics[scale=0.64]{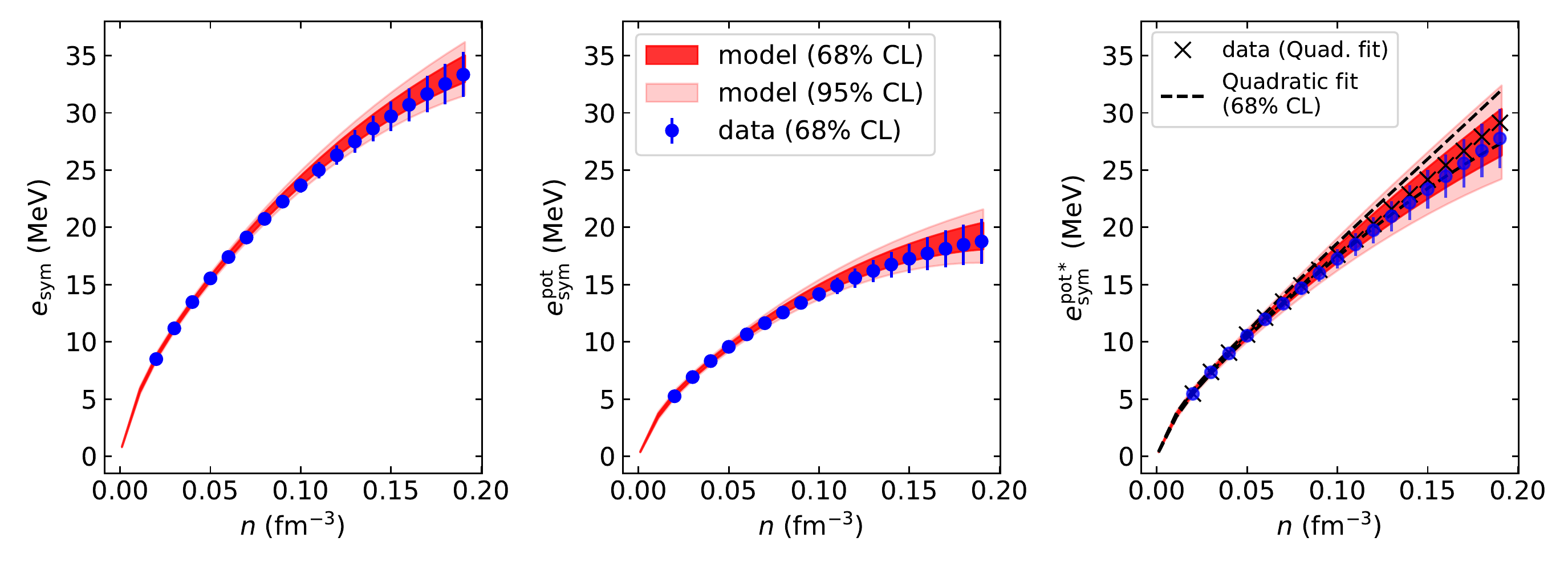}
\caption{Results for the symmetry energy, $e_{\textrm{sym}}(n)$, its potential contribution $e^{\textrm{pot}}_{\textrm{sym}}$, and the effective potential $e^{\textrm{pot}*}_{sym}$ using Eqs.~(\ref{eq:esymtot}), (\ref{eq:e_sym_pot}), and (\ref{eq:e_sym_pot*}).
The meaning of the individual bands and points is the same as in Fig.~\ref{fig:scalings}.
In the right panel, the light- and dark-shaded red bands and the data (blue points) correspond to calculations where the Landau mass is represented by a linear polynomial.
The black squares (without error bars) and the black dashed lines (enclosing a band) represent calculations where the Landau mass is represented by a quadratic fit.}
\label{fig:e_sym}
\end{figure*}

The global symmetry energy $e_{\textrm{sym}}$ is simply determined from the fits of PNM and SNM, see Eq.~\eqref{eq:esym1} and Sec.~\ref{sec:SNMPNM}, while
the contributions of the potential energies $e^{\textrm{pot}}_{\textrm{sym}}$, and $e^{\textrm{pot}*}_{\textrm{sym}}$ are obtained from $e_{\textrm{sym}}$ according to the following definitions,
\begin{eqnarray}
e^{\textrm{pot}}_{\textrm{sym}}(n) &=& e_{\textrm{sym}}(n) - t_{\textrm{PNM}}(n) + t_{\textrm{SNM}}(n) \, ,\label{eq:e_sym_pot}\\
e^{\textrm{pot}*}_{\textrm{sym}}(n) &=& e_{\textrm{sym}}(n) - t^*_{\textrm{PNM}}(n) + t^*_{\textrm{SNM}}(n) \, .
\label{eq:e_sym_pot*}
\end{eqnarray}
We use the fits of the Landau mass, see Sec.~\ref{sec:effmass}, to determine $t^*$, including its uncertainties, as explained in the following. 

We present these quantities in Fig.~\ref{fig:e_sym} as functions of the density $n$.
For $e_{\textrm{sym}}$, the data points are obtained from the PNM and SNM data, while the error bars are defined as the arithmetic average of the PNM and SNM error bars.
For the model, we employ the symmetry energy determined from the MM, which is defined as
\begin{eqnarray}
e_{\textrm{sym}}^\mathrm{MM}(n) = e_{\textrm{PNM}}^\mathrm{MM}(n) - e_{\textrm{SNM}}^\mathrm{MM}(n) \, .
\label{eq:esymtot}
\end{eqnarray}
The results shown in Fig.~\ref{fig:e_sym} are obtained from the best fits to SNM and PNM (scaling~$3*$ in Table~\ref{tab:tbl_2}), where the width of the bands is defined as the arithmetic average of the widths in SNM and PNM.
The model results, therefore, depend on the choice of prior in SNM and PNM, in particular, on the prior knowledge of the saturation density and energy considered in SNM, see the discussion of Fig.~\ref{fig:scalings}. 
For this reason, the MM uncertainty for the symmetry energy is slightly smaller than the uncertainty of the data in Fig.~\ref{fig:e_sym}.
At nuclear saturation density, $n_\mathrm{sat}=0.161(7)\fmiq$, the data suggest  $e_{\textrm{sym}}=30.70(140)$~MeV, while the MM leads to $e_{\textrm{sym}}=31.33(106)$~MeV.
Our values for the symmetry energy are in good agreement with the fiducial value of $31.6\pm2.7$~MeV in Ref.~\cite{Li:2019xxz}, with the recent Bayesian analysis in Refs.~\cite{Drischler:2020a, Drischler:2020b} that fully quantifies correlated EFT truncation errors with chiral NN and 3N interactions up to \NNNLO, 
$30.9(11)$~MeV at the canonical saturation density, 
and with the value of $30(3)$~MeV in Ref.~\cite{Lonardoni:2019ypg}  ($E\mathbbm1$).
Similarly, we predict $L_\mathrm{sym}=46.2(49)$~MeV, while $L_\mathrm{sym}=58.9(160)$~MeV was found in Ref.~\cite{Li:2019xxz}, and $L_\mathrm{sym}=58.4(48)$~MeV in
Refs.~\cite{Drischler:2020a, Drischler:2020b} at the canonical saturation density.
The determination of $L_\mathrm{sym}$, however, is very sensitive to the densities at which the value is extracted as well as to the interactions employed, see, e.g., Ref.~\cite{Lonardoni:2019ypg}.

The data for $e_{\textrm{sym}}^{\textrm{pot}}$ and $e^{\textrm{pot}*}_{\textrm{sym}}$ are obtained from $e_{\textrm{sym}}$ using Eqs.~(\ref{eq:e_sym_pot}) and~(\ref{eq:e_sym_pot*}).
In the case of $e^{\textrm{pot}*}_{\textrm{sym}}$, the effective mass is fixed to be the best fit using either the linear or the quadratic density expansion (depicted by dashed lines in the right panel), and the uncertainty of $e^{\textrm{pot}*}_{\textrm{sym}}$ is defined as the arithmetic average of the uncertainties of $e_{\textrm{sym}}$, $t^*_{\textrm{PNM}}$ and $t^*_{\textrm{SNM}}$.
Therefore, the uncertainty of $e^{\textrm{pot}*}_{\textrm{sym}}$ also includes the uncertainties in the Landau mass parameters $\kappa_{\textrm{sat}}$ and $\kappa_{\textrm{PNM}}$.
We observe that there is a large impact of the Landau mass on $e^{\textrm{pot}*}_{\textrm{sym}}$, compared to $e_{\textrm{sym}}^{\rm pot}$ with the bare mass.
At $n_\textrm{sat}$, we obtain $e_{\textrm{sym}}^{\rm pot}=18.1(7)$~MeV and $e^{\textrm{pot}*}_{\textrm{sym}}=25.7(14)$~MeV.
Hence, the Landau mass increases the potential part of the symmetry energy by about $30-40$\% as discussed in the introduction.
These numbers are compatible with the expectations for the complementary contribution from the kinetic energy.
We find $t_{\textrm{PNM}}(n_{\textrm{sat}}^\mathrm{emp})-t_{\textrm{SNM}}(n_{\textrm{sat}}^\mathrm{emp}) = 13.0$~MeV and $t_{\textrm{PNM}}^*(n_{\textrm{sat}}^\mathrm{emp})-t_{\textrm{SNM}}^*(n_{\textrm{sat}}^\mathrm{emp}) = 5.4(13)$~MeV.

For $e^{\textrm{pot}*}_{\textrm{sym}}$, we expect a difference when using either a Landau mass that is linear or quadratic in density, see Fig.~\ref{fig:ms:fit}.
In Fig.~\ref{fig:e_sym} we show two results for $e^{\textrm{pot}*}_{\textrm{sym}}$.
The blue data points and the dark-shaded (light-shaded) red bands correspond to the results at 68\% (95\%) confidence level when using a Landau mass linear in density.
The black squares and the black-dashed lines, encompassing the corresponding 68\% confidence interval, represent calculations with a Landau mass quadratic in density.
Interestingly, the values for $e^{\textrm{pot}*}_{\textrm{sym}}$ obtained from these two functions for the Landau mass differ only by about $1.8\%$, which is quite small. 
We, therefore, use only the linear fit for the Landau mass in the following.

\subsection{Quadratic contribution to the symmetry energy}

The quadratic contribution to the symmetry energy, $e_{\textrm{sym},2}$, is defined in Eq.~(\ref{eq:esym2}) as the local curvature in the isospin-asymmetry parameter $\delta$ in SNM.
In the following, we extract $e_{\textrm{sym},2}$ using this expansion around SNM, but also suggest obtaining $e_{\textrm{sym},2}$ from an expansion around PNM. 
We demonstrate that both definitions provide comparable results.

\subsubsection{Expansion around SNM}

\begin{figure*}
    \centering
    \includegraphics[scale=0.47]{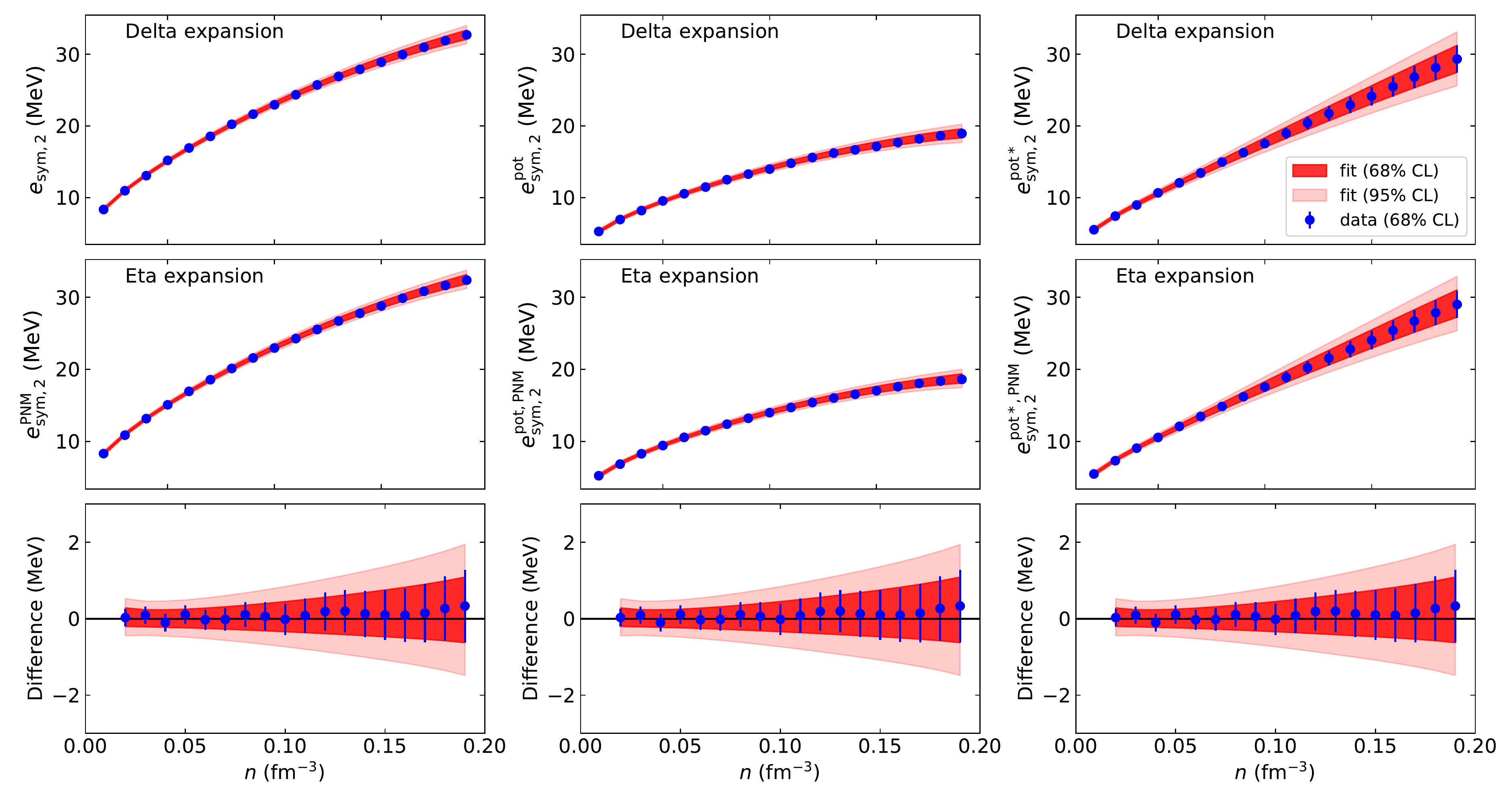}
    \caption{Comparison of the extracted $e_{\textrm{sym},2}$, $e^{\textrm{pot}}_{\textrm{sym},2}$, and $e^{\textrm{pot}*}_{\textrm{sym},2}$ using Eqs.~(\ref{eq:esym2:fit}), (\ref{eq:e_sym2_pot}), and~(\ref{eq:e_sym2_pot*}) (first row) or via an expansion around PNM, i.e., using Eq.~(\ref{eq:esym2_eta}) (second row).
    The third row shows the difference between the $\delta$ and $\eta$ expansions.}
    \label{fig:eta_e_sym2}
\end{figure*}

The quadratic contribution to the symmetry energy is defined by Eq.~(\ref{eq:esym2}) relative to the SNM EOS.
To determine this contribution directly from the MBPT data, we employ Eq.~\eqref{eq:EA_quadratic_expansion} up to the fourth order in $\delta$, and fit the coefficients $e_{\textrm{sym},2}$ and $e_{\textrm{sym},4}$ using \textsf{optimize.curve\_fit} from the Python package \textsc{SciPy}.
The fits are performed for isospin asymmetries in the range, $\delta = 0.0-0.5$. 
We have checked that the results are insensitive (within variations of about $0.1$) to the upper limit of this range---as long as it is chosen to be $\geqslant 0.5$.

For the model, we express $e_{\textrm{sym},2}(n)$ as a function of the density using the MM contribution to the quadratic symmetry energy,
\begin{eqnarray}
e_{\textrm{sym},2}^\mathrm{MM}(x) &=& \frac{5}{9} t_{\textrm{SNM}}(x)  + \sum_{j=0}^{N} \frac{ x^j}{j !} \bigg[ v_{\textrm{sym2},j} u_j(x,\delta=0) - \nonumber \\ && v_{\textrm{SNM,j}} \big(u_j(x,\delta=0) - 1 \big) (1+3x) b_{\textrm{sym}} \bigg]  \, ,
\label{eq:esym2:fit}
\end{eqnarray}
where the parameters $v_{\textrm{sym2},i}$ are determined using Lepage's Python package \lsqfit, as before for other quantities.
They are related to the quadratic symmetry energy NEPs as follows:
\begin{widetext}
\begin{eqnarray}
v_{\textrm{sym2},0} &=& E_{\textrm{sym2}} - \frac{5}{9} t^{\textrm{sat}}_{\textrm{SNM}} [1+\kappa_{\textrm{sat}} + 3\kappa_{\textrm{sym}} +\kappa_{\textrm{sat,2}}   ] \,, \\
v_{\textrm{sym2},1} &=& L_{\textrm{sym2}}- \frac{5}{9} t^{\textrm{sat}}_{\textrm{SNM}} [2+5(\kappa_{\textrm{sat}} + 3\kappa_{\textrm{sym}} ) +8\kappa_{\textrm{sat,2}}   ] \,, \nonumber  \\
v_{\textrm{sym2},2} &=& K_{\textrm{sym2}} - \frac{10}{9} t^{\textrm{sat}}_{\textrm{SNM}}[-1+5(\kappa_{\textrm{sat}} + 3\kappa_{\textrm{sym}}) +20\kappa_{\textrm{sat,2}}   ] \,, \nonumber  \\
v_{\textrm{sym2},3} &=& Q_{\textrm{sym2}} - \frac{10}{9} t^{\textrm{sat}}_{\textrm{SNM}}[4-5(\kappa_{\textrm{sat}} + 3\kappa_{\textrm{sym}}) +40\kappa_{\textrm{sat,2}}   ] \,, \nonumber  \\
v_{\textrm{sym2},4} &=& Z_{\textrm{sym2}} - \frac{40}{9} t^{\textrm{sat}}_{\textrm{SNM}}[-7+5(\kappa_{\textrm{sat}} + 3\kappa_{\textrm{sym}} ) -10\kappa_{\textrm{sat,2}}   ] \,.\nonumber
\end{eqnarray}
\end{widetext}
These relations generalize the ones in Ref.~\cite{Margueron2018a} for a quadratic density-dependent Landau mass.
In Eq.~(\ref{eq:esym2:fit}), the parameter $b_{\textrm{sym}} \equiv b_{\textrm{PNM}} - b_{\textrm{SNM}}$ is fixed by the $3^*$ fit.

The contributions to the symmetry energy due to the potential energy are determined from the following expressions,
\begin{eqnarray}
e^{\textrm{pot}}_{\textrm{sym},2}(n) &=& e_{\textrm{sym},2}(n) - \frac{5}{9} t_{\textrm{SNM}}(n)\, ,\label{eq:e_sym2_pot} \\
e^{\textrm{pot}*}_{\textrm{sym},2}(n) &=& e_{\textrm{sym},2}(n) - \frac{5}{9} t_{\textrm{SNM}}(n)   \bigg[1 +   \kappa_{\textrm{sat}} \frac{n}{n_{\textrm{sat}}} \nonumber \\
&& + \kappa_{\textrm{sat},2} \bigg( \frac{n}{n_{\textrm{sat}}} \bigg)^2 + 3\kappa_{\textrm{sym}} \frac{n}{n_{\textrm{sat}}}\bigg]\, .
\label{eq:e_sym2_pot*}
\end{eqnarray}

Our results for $e_{\textrm{sym},2}$, $e^{\textrm{pot}}_{\textrm{sym},2}$, and $e^{\textrm{pot}*}_{\textrm{sym},2}$ are shown in the first row of Fig.~\ref{fig:eta_e_sym2}.
At $n_\mathrm{sat}^\mathrm{emp}$, we find $e_{\textrm{sym},2}(n_\mathrm{sat}^\mathrm{emp})=30.0(4)$~MeV, $e^{\textrm{pot}}_{\textrm{sym},2}(n_\mathrm{sat}^\mathrm{emp})=17.7(4)$~MeV and $e^{\textrm{pot}*}_{\textrm{sym},2}(n_\mathrm{sat}^\mathrm{emp})=26.4(1.7)$~MeV (with the linear density-dependent model for the Landau mass).
The large value of $e^{\textrm{pot}*}_{\textrm{sym},2}(n_\mathrm{sat}^\mathrm{emp})$, almost 90\% of $e_{\textrm{sym},2}(n_\mathrm{sat}^\mathrm{emp})$ originates from the isospin dependence of the Landau mass, encoded by $\kappa_{\textrm{sym}}$.

From the fit model~(\ref{eq:esym2:fit}), we obtain an estimate for the NEPs that govern the quadratic contribution to the symmetry energy at the inferred value of $n_{\rm sat}$. The values are given in the first row of Table~\ref{tab:NEP2}.
Our result for $E_{\textrm{sym},2}$ is about 1~MeV below the total symmetry energy, $E_{\textrm{sym}}$---the difference is due to non-quadratic contributions.

\begin{table*}
\centering
\tabcolsep=0.4cm
\def\arraystretch{1.9}
\caption{Posteriors of empirical parameters obtained from the analysis of the $\delta$ and $\eta$ expansions for $e_{\textrm{sym},2}$.
Values inside parentheses represent error bars at the $\pm 1\sigma$ level. Results for the following three Skyrme-type interactions are given: NRAPR~\cite{NRAPR}, LNS5~\cite{LNS5} and SAMI~\cite{SAMI}.}
\label{tab:NEP2}
\begin{tabular}{cccccccc}
\hline
 Expansion & $n_{\textrm{sat}}$ & $E_{\textrm{sym},2}$ & $L_{\textrm{sym},2}$ & $K_{\textrm{sym},2}$ & $Q_{\textrm{sym},2}$ & $Z_{\textrm{sym},2}$ & $b_{\textrm{sym}}$ \\
 & (fm$^{-3}$) & (MeV) & (MeV) & (MeV) & (MeV) & (MeV) &  \\
\hline
Prior & - & $31.50(350)$ & $50(10)$ & $-130 (110)$ & $-300(600)$ & $-1800(800)$ \\
\hline
 $\delta$ (SNM) & $0.161(7)$$\dagger$ & $30.16(83)$ & $47(3)$ & $-146(43)$ & $90(334)$ & $-1865(793)$ & $25$$\dagger \dagger$  \\
 $\eta$ (PNM) & $0.161(7)$$\dagger$ & $30.02(82)$ & $46(3)$ & $-149(46)$ & $93(352)$ &  $-1875(793)$ & $25$$\dagger \dagger$ \\
 \hline
 NRAPR~\cite{NRAPR} & $0.161$ & $32.78$ & $60$ & $-123$ & $312$ & $-1836$ \\
 LNS5~\cite{LNS5}   & $0.160$ & $29.15$ & $51$ & $-119$ & $286$ & $-1672$ \\
 SAMI~\cite{SAMI}   & $0.159$ & $28.16$ & $44$ & $-120$ & $372$ & $-2180$ \\
\hline
\end{tabular}

$\dagger$ Priors taken from the SNM posteriors in Table~\ref{tab:tbl_2}. $\dagger \dagger$ Fixed value. 
\end{table*}

\subsubsection{Expansion around PNM}

An alternative approach is to determine the contribution $e_{\textrm{sym},2}$ from an expansion around PNM.
Since the MBPT approach used here is able to explore asymmetric matter with arbitrary $\delta$, we can test the accuracy of this alternative expansion.

To this end, we introduce the parameter
\begin{equation}
    \eta=1-\delta=2n_p/n\,,
\end{equation}
which is twice the proton fraction.
Equation~(\ref{eq:EA_quadratic_expansion}) can now be re-expressed in terms of the this parameter,
\begin{eqnarray}
e(\eta) &=& e_{\textrm{PNM}} -2(e_{\textrm{sym},2}+2e_{\textrm{sym},4}) \eta 
+(e_{\textrm{sym},2}+6e_{\textrm{sym},4}) \eta^2 \nonumber \\
&&-4 e_{\textrm{sym},4} \eta^3 + e_{\textrm{sym},4} \eta^4 + O(\eta^5) \, .
\label{eq:eta:1}
\end{eqnarray}
From Eq.~(\ref{eq:eta:1}), it follows then
\begin{equation}
e_{\textrm{sym},2}^{\textrm{PNM}}(n) = -\frac 3 4 \frac{\partial e}{\partial \eta} \biggr \rvert_{\eta=0} -\frac 1 4 \frac{\partial^2 e}{\partial \eta^2}\biggr\rvert_{\eta=0} \,.
\label{eq:esym:PNM}
\end{equation}

We determine $e_{\textrm{sym},2}^\mathrm{PNM}$ and $e_{\textrm{sym},4}^\mathrm{PNM}$ from fitting the function
\begin{eqnarray}
e(n,\eta) &=& e_{\textrm{PNM}}(n) +a_1 (n) \eta +a_2(n) \eta^2 +\mathcal{O}(\eta^3)\,.
\label{eq:eta:2}
\end{eqnarray}
with
\begin{eqnarray}
e_{\textrm{sym},2}^\mathrm{PNM}(n) &=& -\frac 1 4 [3 a_1(n) + 2 a_2(n)] \, , \label{eq:esym2_eta}\\
e_{\textrm{sym},4}^\mathrm{PNM}(n) &=& \frac 1 8 [a_1(n) + 2 a_2(n)] \, ,
\label{eq:esym4_eta}
\end{eqnarray}
at each density to the computed energies per particle at $\eta =0.0, 0.1, 0.2, \text{and~} 0.3$.
Again, we also perform a Bayesian fit using Eq.~(\ref{eq:esym2:fit}).
The two quantities are shown in the second row of Fig.~\ref{fig:eta_e_sym2}, together with the potential terms $e_{\textrm{sym},2}^{\textrm{pot},\textrm{PNM}}$ and $e_{\textrm{sym},2}^{\textrm{pot}*,\textrm{PNM}}$.
The NEPs obtained from Eq.~(\ref{eq:esym2:fit}) are given in the second row of Table~\ref{tab:NEP2}.
Note, that the differences between the NEPs extracted around SNM [using Eq.~(\ref{eq:esym2})] and around PNM [using Eq.~\eqref{eq:esym:PNM}] are much smaller than the uncertainties of these NEPs, which demonstrates that the two approaches are consistent with one another.
This is further illustrated in the third row of Fig.~\ref{fig:eta_e_sym2}, where the difference between $e_{\textrm{sym},2}^\mathrm{SNM}$
and $e_{\textrm{sym},2}^\mathrm{PNM}$ is shown to be consistent with zero and a small width of about 1.5~MeV at $\nsat$.

Determining the quadratic contribution to the symmetry energy from an expansion around PNM might be beneficial because PNM can usually be computed with much higher accuracy since the uncertainties in the 3N interactions are reduced.
Furthermore, such an extraction is useful for microscopic approaches in which a small proton impurity can be treated more easily than SNM, e.g., the auxiliary-field diffusion Monte Carlo approach~\cite{Lonardoni:2019ypg}.

\subsection{Non-quadratic contribution to the symmetry energy $e_{\textrm{sym},nq}$ and $e_{\textrm{sym},4}$}
\label{sec:nqesym}

\begin{figure*}
\centering
\includegraphics[scale=0.64]{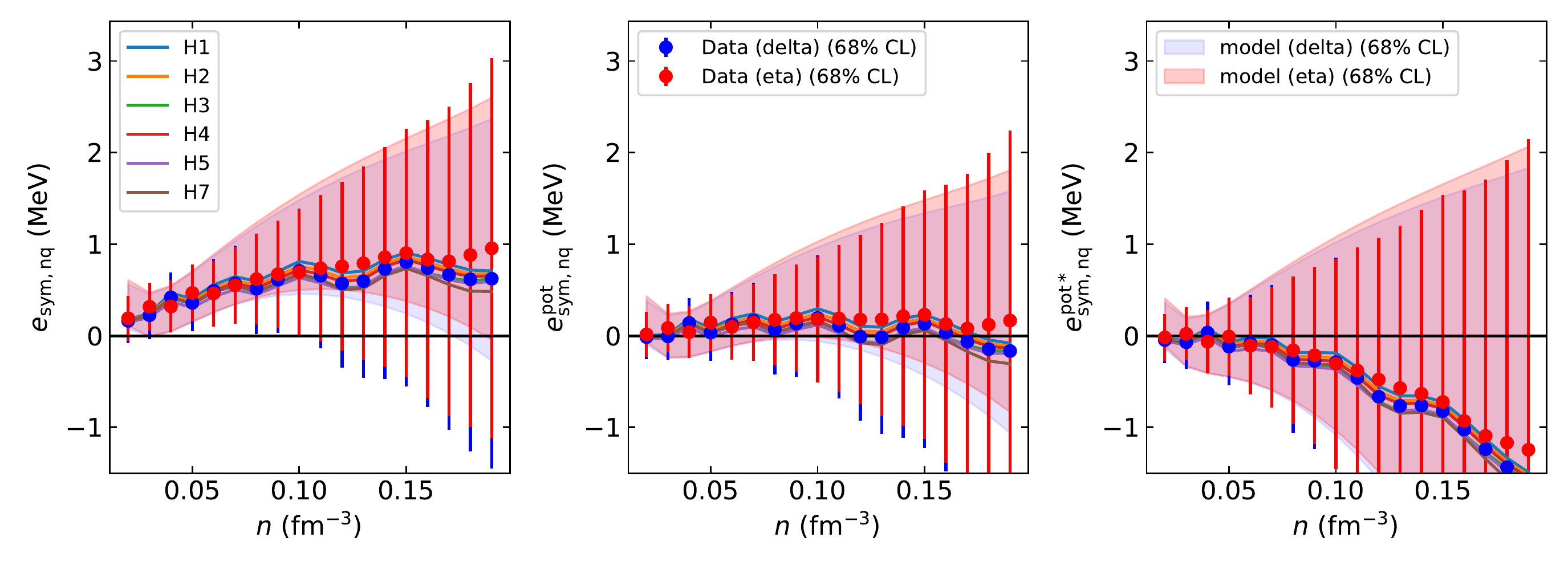}
\caption{Non-quadratic terms $e_{\textrm{sym},nq}$ calculated via an expansion around SNM (blue) and PNM (red). 
For the right panel, the Landau mass is described by a linear fit. 
The coloured lines depict results for the six individual Hamiltonians. 
}
\label{fig:e_symnq}
\end{figure*}

We now evaluate the contribution of the non-quadratic terms, defined by Eq.~\eqref{eq:general:symnq}, using the expansions around SNM and PNM, respectively.
For the global symmetry energy, we use our model~\eqref{eq:esymtot}, while for describing the quadratic contribution we use the fit~\eqref{eq:esym2:fit}. 
Figure~\ref{fig:e_symnq} shows our results for the expansion around SNM (blue) and the expansion around PNM (red). 
Both expansions agree, and the differences are smaller than the uncertainties by a factor of $2-3$.
We also show results for the six Hamiltonians. 
Their spread is much smaller than the uncertainties of the data or the model. This is because the latter are computed as arithmetic averages of the error bars of the global symmetry energy and the quadratic contribution.
At $n_{\textrm{sat}}^\mathrm{emp}$,  we obtain from our model $e_{\textrm{sym},nq}=1.3(10)$~MeV, $e_{\textrm{sym},nq}^{pot}=0.6(10)$~MeV, $e_{\textrm{sym},nq}^{pot*}=-0.5(22)$~MeV and from the data, $e_{\textrm{sym},nq}=0.8(15)$~MeV, $e_{\textrm{sym},nq}^{pot}=0.1(15)$~MeV, and $e_{\textrm{sym},nq}^{pot*}=-0.9(25)$~MeV .
We find that these non-quadratic contributions represent a correction of about $3-5$\% to the symmetry energy. They originate mainly from the kinetic energy, since $e_{\textrm{sym},nq}^{pot}$ remains close to zero across all densities.
The associated uncertainties represent the model dependence explored in the present approach. 
Our estimates for the non-quadratic contributions to the symmetry energy and the NEP are summarized in Table~\ref{tab:NEPNQ}.
We also compare the present NEPs to the three selected Skyrme interactions, showing a good agreement between the microscopic results and the EDF approaches.

\begin{table*}
\centering
\tabcolsep=0.7cm
\def\arraystretch{1.9}
\caption{Posteriors of non-quadratic and quartic NEPs obtained from the analysis of $e_{\textrm{sym},nq}$ using the $\delta$ and $\eta$ expansions, and $e_{\textrm{sym},4}$ obtained from the $\eta$ expansion only.
For the extraction of $e_{\textrm{sym},nq}$ via the $\eta$ expansion, the values inside the square brackets are obtained from a fit to the data in order to provide a direct comparison to the corresponding analysis of $e_{\textrm{sym},4}$.
Values in parenthesis represent the $\pm 1\sigma$ uncertainties.
The NEPs for the following three Skyrme-type interactions are given: NRAPR~\cite{NRAPR}, LNS5~\cite{LNS5} and SAMI~\cite{SAMI}.}
\label{tab:NEPNQ}
\begin{tabular}{lccccc}
\hline
 Non-quadratic & $E_{\textrm{sym},nq}$ & $L_{\textrm{sym},nq}$ & $K_{\textrm{sym},nq}$ & $Q_{\textrm{sym},nq}$ & $Z_{\textrm{sym},nq}$ \\
 contribution & (MeV) & (MeV) & (MeV) & (MeV) & (MeV)  \\
\hline
 SNM &  $1.2(15)$ & $0(6)$ & $-24(58)$ &  $106(426)$ & $ 91(1057)$ \\
 PNM &   $1.3(15)$ &  $1(6)$ &  $-20(60)$ &  $103(441)$ & $101(1058)$  \\
   &   $\big[ 0.84(7)  \big]$ &  $\big[ 0.7(8)  \big]$ &  $\big[ -9(13)  \big]$ &  $\big[ 32(151)  \big]$ & $\big[  167(958)  \big]$  \\
\hline
NRAPR~\cite{NRAPR} & $1.40$ & $5$ & $6$ & $-1$ &  $-12$ \\
LNS5~\cite{LNS5}   & $1.70$ & $6$ & $9$ & $-4$ &  $1$ \\
SAMI~\cite{SAMI}   & $1.08$ & $3$ & $2$ & $2$ & $-24$ \\
\hline
\hline
 Quartic & $E_{\textrm{sym},4}$ & $L_{\textrm{sym},4}$ & $K_{\textrm{sym},4}$ & $Q_{\textrm{sym},4}$ & $Z_{\textrm{sym},4}$ \\
 contribution & (MeV) & (MeV) & (MeV) & (MeV) & (MeV)  \\
\hline
 PNM  &  $1.00(8)$ &  $ 0.6(6)$ &  $-7(12)$ &  $69(145)$ &  $ 33(956)$ \\
\hline
NRAPR~\cite{NRAPR} & $0.95$  & $3$  & $4$ & $-1$ &  $-6$  \\
LNS5~\cite{LNS5}   & $1.17$  & $5$  & $6$ & $-4$ &  $3$  \\
SAMI~\cite{SAMI}   & $0.70$  & $2$  & $2$ &  $1$ & $-15$  \\
\hline
\end{tabular}
\end{table*}

We calculate the quartic contribution to the symmetry energy $e_{\textrm{sym},4}$ using Eq.~\eqref{eq:esym4_eta}, and show the resulting NEPs in Table~\ref{tab:NEPNQ}.
We find that the quartic term to the symmetry energy accounts for about $60-70$\% of the total non-quadratic contribution, while the remaining $30-40$\% originate from higher-order contributions.
The convergence of these additional contributions is discussed in Ref.~\cite{Wellenhofer2016}.
We stress that this does not include any logarithmic contribution because such a contribution would vanish in PNM.

A recent analysis based on a general EDF approach---which was optimized to the properties of finite nuclei---concluded that quartic terms $\propto \delta^4$ have little impact on nuclei~\cite{Bulgac2018}. 
The result was interpreted as a consequence of the fact that the asymmetry $\delta$ in finite nuclei is small: for $Z>8$ it varies between $-0.33$ and $+0.38$ in the latest Atomic Mass Data Center mass table AME2016~\cite{AME2016}.
A quartic term was, however, found to be important to correctly reproduce the PNM energy per particle.
In order to reproduce the PNM energy per particle predicted in Ref.~\cite{Wlazlowski2014}, Ref.~\cite{Bulgac2018} found a quartic term of $e_{\textrm{sym},4}= 2.635 \MeV$at $n=0.1 \fmiq$.
This term was the only non-quadratic contribution considered in Ref.~\cite{Bulgac2018}, and is consistent within our upper 68\% confidence interval for the non-quadratic contribution to the symmetry energy.
The higher value for $e_{\textrm{sym},4}$ obtained in Ref.~\cite{Bulgac2018} might be related to the larger value in the PNM energy per particle obtained in Ref.~\cite{Wlazlowski2014}, as shown in Fig.~\ref{fig:PNM_3_add}.
This affects the symmetry energy because the contribution $e_{\textrm{sym},4}$ is needed to correctly reproduce the PNM EOS.
Both, the PNM energy per particle in Ref.~\cite{Wlazlowski2014} and $e_{\textrm{sym},4}$ obtained in Ref.~\cite{Bulgac2018} are $\sim 1$~MeV higher than the values we obtain in the this work.

\subsection{Logarithmic contribution to the symmetry energy $e_{\textrm{\textrm{sym},log}}$}

\begin{figure*}
\centering
\includegraphics[scale=0.47]{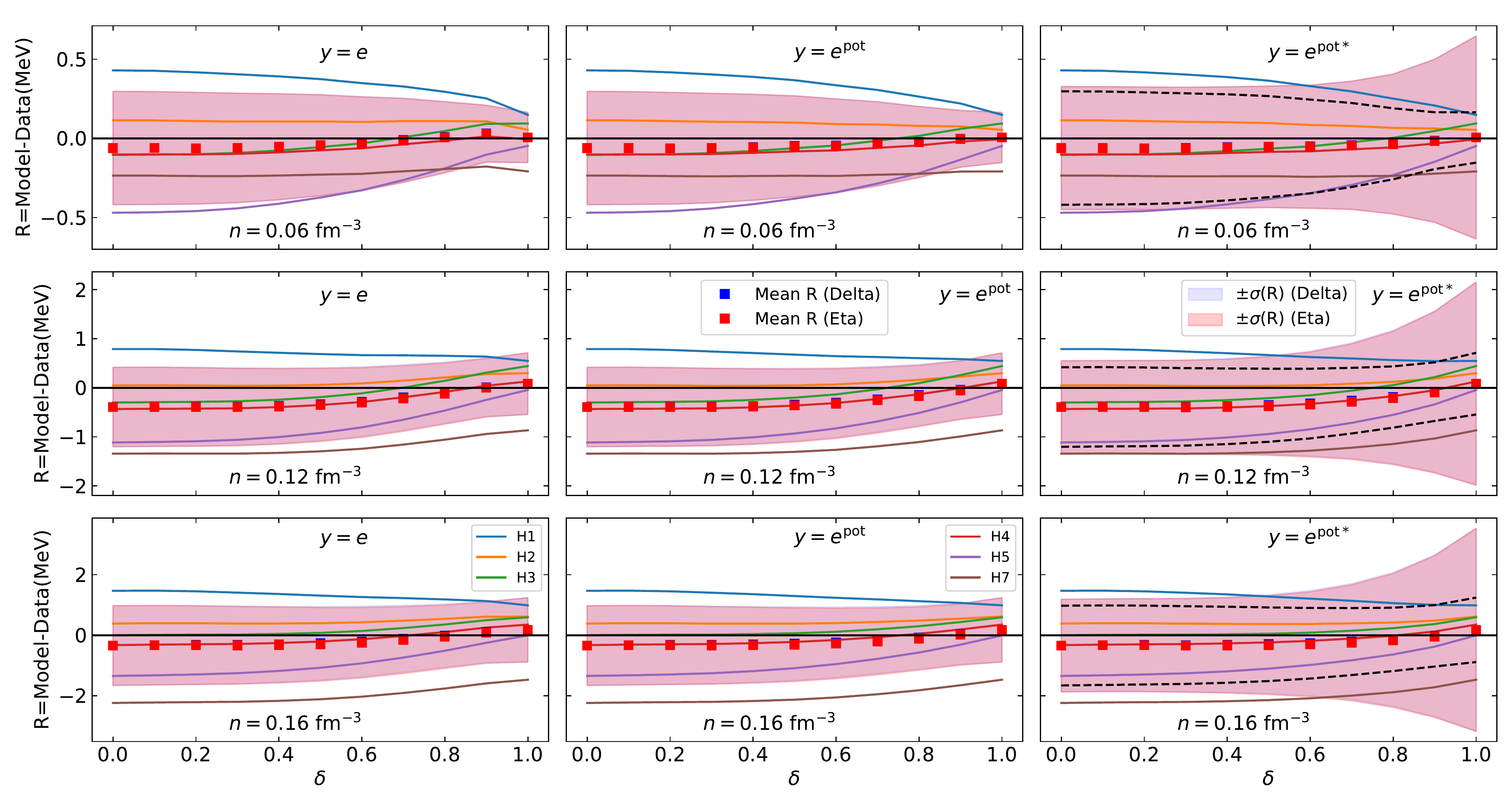}
\caption{Residuals of the model, see Eq. (\ref{eq:exp1}), with respect to the data as a function of the asymmetry parameter $\delta$ for different values of the density: 
$n=0.06$~fm$^{-3}$ (first row), 
$n=0.12$~fm$^{-3}$ (second row), 
and $n=0.16$~fm$^{-3}$ (third row).
The results are shown for the two different calculations of $y_{\textrm{\textrm{sym},2}}$ and $y_{\textrm{\textrm{sym},nq}}$---the expansions around PNM (red) and SNM (blue).
The different columns correspond to $e$, $e^{\textrm{pot}}$, and $e^{\textrm{pot},*}$.
The coloured lines depict the residuals of the fit for each Hamiltonian. 
In the last column, the black-dashed lines represent the upper and lower limits of the uncertainty in the residuals, respectively, by disregarding the uncertainties in the effective masses.}
\label{fig:residues}
\end{figure*}

The leading-order logarithmic contribution to the symmetry energy, see Eq.~(\ref{eq:general}), was suggested to be of the form $\delta^4\log\vert\delta\vert$~\cite{Kaiser2015, Wellenhofer2016}.
It, therefore, vanishes in both SNM and PNM, and data at finite isospin asymmetry are required to determine its magnitude.
Such a logarithmic term would appear by a characteristic arch-like structure in the $\delta$-dependent residuals between the data and a model without the logarithmic term.
Figure~\ref{fig:residues} shows these residuals as a function of $\delta$ at three different densities.
For asymmetric matter,  we use
\begin{align}
y^{\textrm{model}}(n,\delta) =
 y_{\textrm{SNM}}(n)+y_{\textrm{sym},2}(n)\delta^2+y_{\textrm{\textrm{sym},nq}}(n)\delta^4 \,,
\label{eq:exp1}
\end{align}
where $e_{\textrm{SNM}}$, $e_{\textrm{sym},2}$ and $e_{\textrm{\textrm{sym},nq}}$
are given by Eqs.~\eqref{eq_fit_SNM}, \eqref{eq:esym2:fit},  and \eqref{eq:general:symnq}.
Note that in the model~\eqref{eq:exp1}, the fourth-order $\delta$ term also includes possible higher-order  contributions (like, for instance, a $\delta^6$ term) contained in the term $y_{\textrm{\textrm{sym},nq}}$.
The different panels in Fig.~\ref{fig:residues} show the results at three densities, $n=0.06$, $0.12$, and $0.16 \fmiq$ (from the top to the bottom panel), and for the three choices for the variable $y$: $e$, $e^{\textrm{pot}}$, and $e^{\textrm{pot*}}$ (from left to right) as a function of the isospin asymmetry $\delta$. 
The squares (shaded bands) represent the mean ($68\%$ confidence level) of the residuals. 
The presence of logarithmic terms would appear as a systematic deviation of the these residuals from zero in asymmetric matter. 
However, this is not what we observe at the three considered densities and for all energy observables.
Instead, we find the residuals to be compatible with zero and almost flat as a function of the isospin asymmetry. 
This is also the case for the results for each Hamiltonian, which we show as coloured lines.
The results for the individual Hamiltonians vanish on average, but the uncertainty bands remain quite sizable, about $\pm 1-2 $~MeV around saturation density. 
Therefore, our findings suggest that there is no statistically significant indication for a net logarithmic contribution to the symmetry energy for the chiral NN and 3N Hamiltonians used in this work. 

Our conclusion about the strength of the logarithmic term is not in contradiction with the findings presented in Refs.~\cite{Kaiser2015,Wellenhofer2016}.
The size of the logarithmic term shown in Ref.~\cite{Wellenhofer2016} is at most $\sim 0.1$~MeV in neutron-rich matter, and only one Hamiltonian was considered.\footnote{This is the n3lo450 interaction constructed in Refs.~\cite{Entem2003,Coraggio2014,Coraggio:2012ca}, which is not considered in this work.} Contributions of the order of $\sim 0.1$~MeV are small compared to the overall theoretical uncertainties, which we estimate by analyzing the six Hamiltonians in Table~\ref{tab:used_hamiltonians}.

\section{Impact on the neutron-star crust-core transition}
\label{sec:SPINO}

In this section,
we study the impact of the non-quadratic contribution to the symmetry energy on the crust-core transition in neutron stars, for which the symmetry energy plays an important role~\cite{Pethick1995, Ducoin2010, Ducoin2011}.
This transition occurs at the core-crust transition density $n_{cc}$ with an isospin asymmetry $\delta_{cc}$ that is determined by the beta-equilibrium. 
The parameters $n_{cc}$ and $\delta_{cc}$ can be obtained from uniform matter by determining the density at which matter becomes unstable with respect to density fluctuations (spinodal instability)~\cite{Pethick1995}.

\begin{figure}
\centering
\includegraphics[scale=0.48]{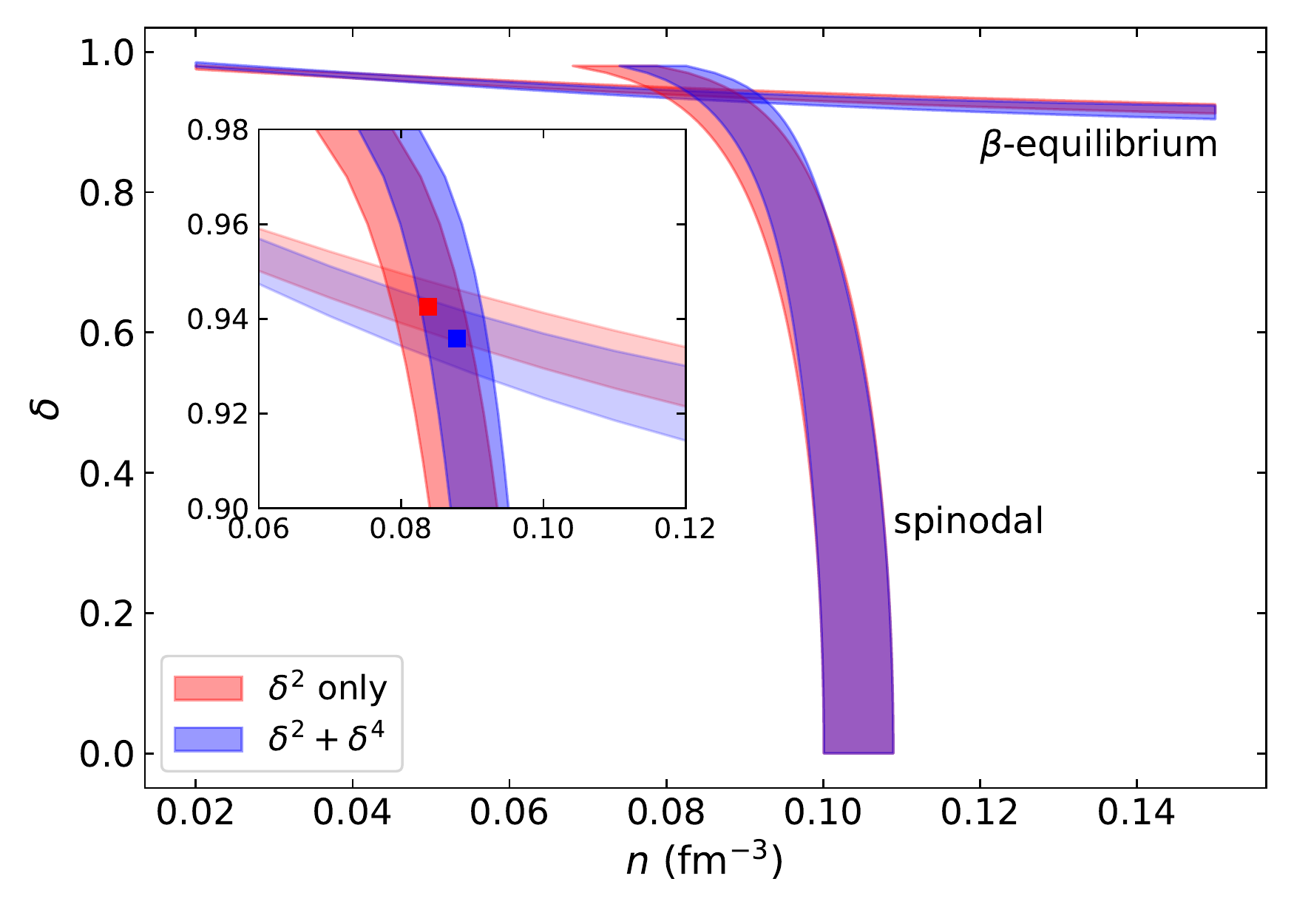}
\caption{Predictions for beta equilibrium in low-density uniform matter obtained by solving Eq.~\eqref{eq:beta}, and for the spinodal density by solving Eq.~\eqref{eq:spino}.
The intersection denotes the crust-core transition, as indicated by a dot in the inset.
The quadratic approximation (red band) is compared to the case where quartic contributions are included (blue band).
}
\label{fig:crust_core}
\end{figure}

In multi-component matter, e.g., matter that consists of neutrons and protons, this spinodal density is determined from the curvature (Hessian) matrix $\mathcal{C}$, defined as the second derivative of the energy density with respect to the component densities~\cite{Pethick1995}.
From the eigenvalues of $\mathcal{C}$, one can determine the stability of matter:
if all eigenvalues are positive, i.e. if $\mathcal{C}$ is positive semi-definite, the matter exhibits a local stability against density fluctuations of all components in any combination, while a change of sign for any eigenvalue triggers an instability with respect to density fluctuations indicated by its associated eigenvector.
The change of sign of the eigenvalues can be extracted from the determinant of $\mathcal{C}$, which reads in nuclear matter,
\begin{equation}
\det \mathcal{C}(n,\delta)= \frac{\partial\mu_n}{\partial n_n} \frac{\partial\mu_p}{\partial n_p} - \frac{\partial\mu_n}{\partial n_p} \frac{\partial\mu_p}{\partial n_n} \, ,
\label{eq:spino}
\end{equation}
where $\mu_n$ and $\mu_p$ are the neutron and proton chemical potentials.
For simplicity, we have neglected the finite-size contribution from the Coulomb interaction as well as the gradient density terms induced by the finite range of nuclear interactions.
It is expected that these terms reduce the spinodal density by only about $0.01$~fm$^{-3}$~\cite{Pethick1995, Ducoin2011}.

In sub-saturation asymmetric matter, the equilibrium state is the state that satisfies the chemical potential equilibrium $\mu_n = \mu_p + \mu_e$, at fixed baryon number $n = n_n + n_p$ and charge neutrality $n_e = n_p$.
At zero temperature, and considering relativistic electrons, this system of equations reduces to a single non-linear equation,
\begin{equation}
\sqrt{m_e^2  +  \left(\frac{3\pi^2}{2} (1-\delta_\beta) n\right)^{2/3}} = 2\frac{ \partial e(n,\delta_\beta)}{\partial\delta} \, ,
\label{eq:beta}
\end{equation}
whose solution, $\delta_\beta(n)$, is obtained by using a combination of the bisection and the secant methods implemented in the Python package of Ref.~\cite{Lepage_gvar}.
Then, we define the crust-core transition as the solution $(n_{cc},\delta_{cc})$ to both, the instability onset criterion, $\det \mathcal{C}=0$, and the beta equilibrium condition, e.g., Eq.~(\ref{eq:beta}).
Equivalently, $n_{cc}$ is defined as the spinodal density in beta equilibrium, where $\delta_{cc}=\delta_\beta(n_{cc})$.
Figure~\ref{fig:crust_core} shows the intersection between these two determinations.
We investigate the purely quadratic approximation for the symmetry energy, with the NEPs given in Table~\ref{tab:NEP2}, and with the quartic terms from Table~\ref{tab:NEPNQ} included.
For all cases, the reference MM in SNM is determined by the best fit given in Table~\ref{tab:tbl_2} for the scaling~$3^*$.

\begin{table}[t]
\centering
\tabcolsep=0.7cm
\def\arraystretch{1.9}
\caption{\label{tab:crust_core} Crust-core transition density and isospin asymmetry, $n_{cc}$ and $\delta_{cc}$, respectively, for the purely quadratic case ($\delta^2$ only) and for the case including the quartic contribution ($\delta^2+\delta^4$), see Eq.~(\ref{eq:exp1}).}
\begin{tabular}{lcc}
\hline
 Model & $n_{cc}$~(fm$^{-3}$)  & $ \delta_{cc}$  \\
\hline
$\delta^2$ only & $0.083(5)$ & $0.944(5)$ \\
$\delta^2+\delta^4$ & $0.087(4)$ & $0.935(6)$ \\
\hline
\end{tabular}
\end{table}

When we include quartic contributions, the spinodal density in neutron-rich matter is increased compared to the case where only the quadratic term is considered. This is because the quartic term increases the symmetry energy.
For the same reason, the isospin asymmetry is decreased when non-quadraticities are included.
Our results are summarized in Table~\ref{tab:crust_core}, and depicted in Fig.~\ref{fig:crust_core} by the blue and red points.
They are in agreement with the predictions of, e.g., Refs.~\cite{Ducoin2010, Ducoin2011} with $L_{\textrm{sym},2}\approx 45$~MeV.
From the comparison of our results with and without the quartic term, we find that the transition density changes by $\sim 5\%$ while $\delta_{cc}$ changes by only $\sim 1\%$.
We conclude that non-quadratic contributions to the symmetry energy have no significant effect on the crust-core transition.

\section{Summary and Conclusions}
\label{sec:CONCLUSIONS}

We have analyzed the properties of asymmetric nuclear matter based on MBPT calculations~\cite{Drischler:2015eba} for six commonly used chiral EFT Hamiltonians with NN and 3N interactions. 
The global symmetry energy, i.e., the difference between EOS in the limits of PNM and SNM, as well as its quadratic and quartic contributions have been determined with theoretical uncertainty estimates.
We have calculated the quadratic contribution to the symmetry energy from the usual expansion around SNM, and have also employed a non-standard approach using an expansion for small proton fractions around PNM. The two approaches are in excellent agreement. Furthermore, we have investigated the strength of the non-quadraticities as well as their model dependence.
The quartic contribution to the symmetry energy was found to be about $1.00(8)$~MeV (or $0.55(8)$~MeV for the potential part).
We have then investigated the leading-order logarithmic term to the symmetry energy, and obtained residuals between our best fit (including quadratic and quartic contributions) and the data to be compatible with zero.
In particular, we found that all residuals where flat in the isospin asymmetry $\delta$, indicating no systematic deviation from zero as expected for a logarithmic contribution. 
However, we also saw that present uncertainties, indicated by the dispersion of the six Hamiltonians of about $1-2$~MeV, are too large to precisely determine its strength. 

Finally, we analyzed the impact of our results on the determination of the crust-core transition in neutron stars using a simple model in the thermodynamic limit. 
We found that the crust-core transition density is increased by $\sim 5\%$, and the associated isospin-asymmetry $\delta$ decreased by $\sim 1\%$ when non-quadraticities are included. 
Hence, these contributions are only small corrections, but need to be included for a precise calculation of the core-crust transition properties.

To gauge the full theoretical uncertainties of the non-quadratic contributions to the symmetry energy, future analyses need to explore a wider range of nuclear interactions and additional asymmetric-matter calculations using different many-body approaches and regularization schemes. In particular, this requires the development of improved chiral NN and 3N interactions up to \NNNLO~\cite{Drischler:2017wtt, Hoppe:2019uyw, Huther:2019ont}, which will enable order-by-order analyses of the neutron-rich matter EOS with statistically meaningful uncertainty estimates derived from chiral EFT~\cite{Drischler:2020a,Drischler:2020b}. 
Our work provides a framework, e.g., Python codes~\cite{rahul_2020_4010747} and Supplemental Material related to our data, for future investigations of the isospin-dependence of nuclear matter.

\acknowledgements

We thank K.~Hebeler, J.~Lattimer, and A.~Schwenk for fruitful discussions.
R.S. is supported by the PHAST doctoral school (ED52) of \textsl{Universit\'e de Lyon}. R.S. and J.M. are both associated to the CNRS/IN2P3 NewMAC project, and are also grateful to PHAROS COST Action MP16214 and to the LABEX Lyon Institute of Origins (ANR-10-LABX-0066) of the \textsl{Universit\'e de Lyon} for its financial support within the program \textsl{Investissements d'Avenir} (ANR-11-IDEX-0007) of the French government operated by the National Research Agency (ANR).
C.D. acknowledges support by the Alexander von Humboldt Foundation through a Feodor-Lynen Fellowship and the US Department of Energy, the Office of Science, the Office of Nuclear Physics, and SciDAC under awards \mbox{DE-SC00046548} and \mbox{DE-AC02-05CH11231}.
The work of I.T. was supported by the U.S. Department of Energy, Office of Science, Office of Nuclear Physics, under contract No.~DE-AC52-06NA25396, by the NUCLEI SciDAC program, and by the LDRD program at LANL.

\appendix

\section{Details of the Fitting Procedure}
\label{fit_procedure}

To obtain the parameter values, we perform in this paper  a Bayesian analysis using the nonlinear least-square fitting  package \textsc{lsqfit} developed in Python~\cite{Lepage_lsqfit}. \textsc{lsqfit} uses \textsc{scipy}'s least-squares minimization routine to optimize our model with respect to data $y_i$ (here, the predictions from each of the six Hamiltonians considered in this work).
The index $i$ runs over all data sampling points (e.g., the density grid).
The propagation of the Gaussian uncertainties from the model parameters, $p_{\alpha}$, to the functions of those parameters, $f(\{p_\alpha\})$, requires calculating derivatives of those functions with respect to the parameters, which is achieved by automatic differentiation.
In this work, the spread of the six Hamiltonians is interpreted as a fair representation of the width $\sigma$ of a normal distribution that reflects theoretical uncertainties.

The fitting procedure requires the minimization of the objective function, hereafter noted the $\chi^2$ function, which, in general, receives contributions from both the input data ($\chi^2_{\text{data}}$) and the prior information on the model parameters ($\chi^2_{\text{prior}}$). We can write these two contributions as
\begin{eqnarray}
\chi^2_{\text{data}}  &=& \sum_{ij} \Delta y(p)_i \  \text{cov}_{ij}^{-1} \ \Delta y(p)_j \label{eq:chi_1}\, ,\\
\chi^2_{\text{prior}} &=& \sum_{\alpha}  \frac{(p_{\alpha}-\mu_{\alpha})^2}{\sigma_{\alpha}^2} \, , \label{eq:chi_2}
\end{eqnarray}
with the total $\chi^2$ being the sum of the two terms,
\begin{equation}
\chi^2=\chi^2_{\text{data}}+\chi^2_{\text{prior}}\,.
\end{equation}
In Eq.~\eqref{eq:chi_1}, $\Delta y(p)_i = f(\{p_\alpha\})_i - E(y_i)$, where the expectation value $E(y_i)$ is defined as
\begin{equation}
E(y_i) = \frac 1 6 \sum_{\mu=1}^{6} (y_i)_{\mu} \, ,
\end{equation}
the summation index $\mu$ runs over the six nuclear Hamiltonians,
and $\text{cov}_{ij}$ is the co-variance matrix between the data points $y_i$ and $y_j$, which is given by
\begin{equation}
\text{cov}_{ij}\equiv\text{cov}(y_i,y_j) = E(y_iy_j)-E(y_i)E(y_j)\, .
\label{eq_corr}
\end{equation}
The correlation matrix $\text{corr}_{ij}$ is then defined by its matrix elements,
\begin{equation}
\text{corr}_{ij} = \frac{\text{cov}_{ij}}{\sqrt{\text{cov}_{ii} \  \text{cov}_{jj}}}\,.
\label{eq:correlation}
\end{equation}

If the data are independent from each other, the co-variance and correlation matrices are diagonal and $\chi^2_{\text{data}}$ is associated with a normal distribution, as it is for $\chi^2_{\text{prior}}$.
Here, the data are not independent because of correlations in density: the knowledge of the predictions of the Hamiltonians at a few density points can be used to determine other points by interpolation or---to some extent---by extrapolation. We, thus, expect a non-diagonal co-variance matrix.
While the Hamiltonians are by construction strongly correlated, we do not account for their correlations in the co-variance matrix.
It only includes the correlations between the different data points, see Eq.~\eqref{eq_corr}.

In Eq.~(\ref{eq:chi_2}), $\mu_{\alpha}$ is the prior mean value of the parameter $p_{\alpha}$, and $\sigma_{\alpha}$ is its standard deviation.
Note that, in this work, we only consider uncorrelated Gaussian prior distributions for our fit parameters.

The best-fit values of the parameters $\bar{p}_{\alpha}$ are those that minimize the total $\chi^2$.
The inverse co-variance matrix corresponding to the posterior distribution of the best-fit parameters is defined as,
\begin{equation}
( \text{cov}^{-1}_p )_{\alpha \beta}  = \frac{\partial \chi^2}{\partial p_{\alpha} \partial p_{\beta}} (\bar{p})\, .
\end{equation}
This expression is used by \lsqfit~to define the uncertainties in the fit parameters $p_{\alpha}$.

\section{Correlations in the data sample}
\label{effective_data}

The data analysis performed in this work involves parametric fits to data that are highly correlated across densities (or equivalently, Fermi momenta). It was found that this correlation had to be taken into account in order to achieve compatibility between the obtained posterior distributions and the data, as shown in Figs. \ref{fig:ms:fit}, \ref{fig:scalings} and \ref{fig:eta_e_sym2}. In this appendix, we provide an estimate of this correlation. 

As mentioned in Appendix~\ref{fit_procedure}, the correlation in the input data is captured by the covaraince matrix~(\ref{eq_corr}). The strong positive correlation in the data points across densities results in large off-diagonal elements with respect to the diagonal ones. We analyse the strength of these off-diagonal elements as follow.  We first diagonalize the correlation matrix~(\ref{eq:correlation}) and obtain the eigenvalues and associated eigenvectors. In the case of maximum correlation all but one eigenvalue are $0$. In our case, we found that the largest eigenvalue contributes to about $96 \%$ of the trace of the correlation matrix. Then, the eigenvector corresponding to this dominant eigenvalue captures the correlation (or mixing) between the data, which needs to be quantified. Let $\lambda$ be this eigenvector with components $(\lambda_1, \dots , \lambda_{N_D} )$, where $N_D$ is the number of data points.  In the case of maximum correlation, the quantities $\vert\lambda_1\vert$, $\dots$, $\vert\lambda_{N_D}\vert$ have zero dispersion about their average value, and in the case of minimum correlation, this dispersion takes a maximum value. We thus define the dispersion about the mean as
\begin{equation}
d^2 = \frac{1}{N_D} \sum_{i=1}^{N_D} (\vert\lambda_i\vert - \bar{\lambda})^2 \,,
\label{eq:dispersion}
\end{equation}
where $\bar{\lambda}$ is the average value of the $\vert\lambda_i\vert$. In the case of a maximum correlation $d^2 = 0$, whereas in the case of minimum correlation $d^2 = \frac{1}{N_D} [1 - \frac{1}{N_D}]$.

We then define the level of correlation as 
\begin{equation}
l_{\rm corr} \equiv 1-N_D\,d^2 \, ,
\label{eq:nheff}
\end{equation}
where $l_{\rm corr} = 1$ in the case of maximum correlation and, for zero correlation $l_{\rm corr} = 1/N_{D}$, which approaches $0$ in the limit of infinitely many data points. This parameter is similar to the correlation coefficient, $\rho$ which one introduces in the case of identical off-diagonal elements in the covariance matrix. 

The values of $l_{\rm corr}$ for the various fits presented in the paper are as follows. In Sec.~\ref{sec:effmass}, fits to the effective masses were presented in Fig.~\ref{fig:ms:fit}. For the linear fit, $l_{\rm corr}$ is $0.96$ in SNM and $0.92$ in PNM; for the quadratic fit, $0.54$ in SNM and $0.54$ in PNM. For the fits shown in Fig.~\ref{fig:scalings}, $l_{\rm corr}$ is $0.91,0.53,0.46$ in the scalings~1, 2 and~3, respectively, in SNM. Similarly in PNM, it is $0.92,0.38,0.39$. Finally, for the fits presented Fig.~\ref{fig:eta_e_sym2}, $l_{\rm corr}=0.50$ for the analysis with the expansion around SNM and $l_{\rm corr}= 0.58$ for the expansion around PNM. 
In all cases, we see that there is a significant correlation in the data. This is, as mentioned before, an important ingredient of our analysis, and is required for agreement between the obtained posterior distributions and the data.

\section{Goodness of the fit and Q--Q plots}

\begin{figure}[t]
\centering
\includegraphics[scale=0.5]{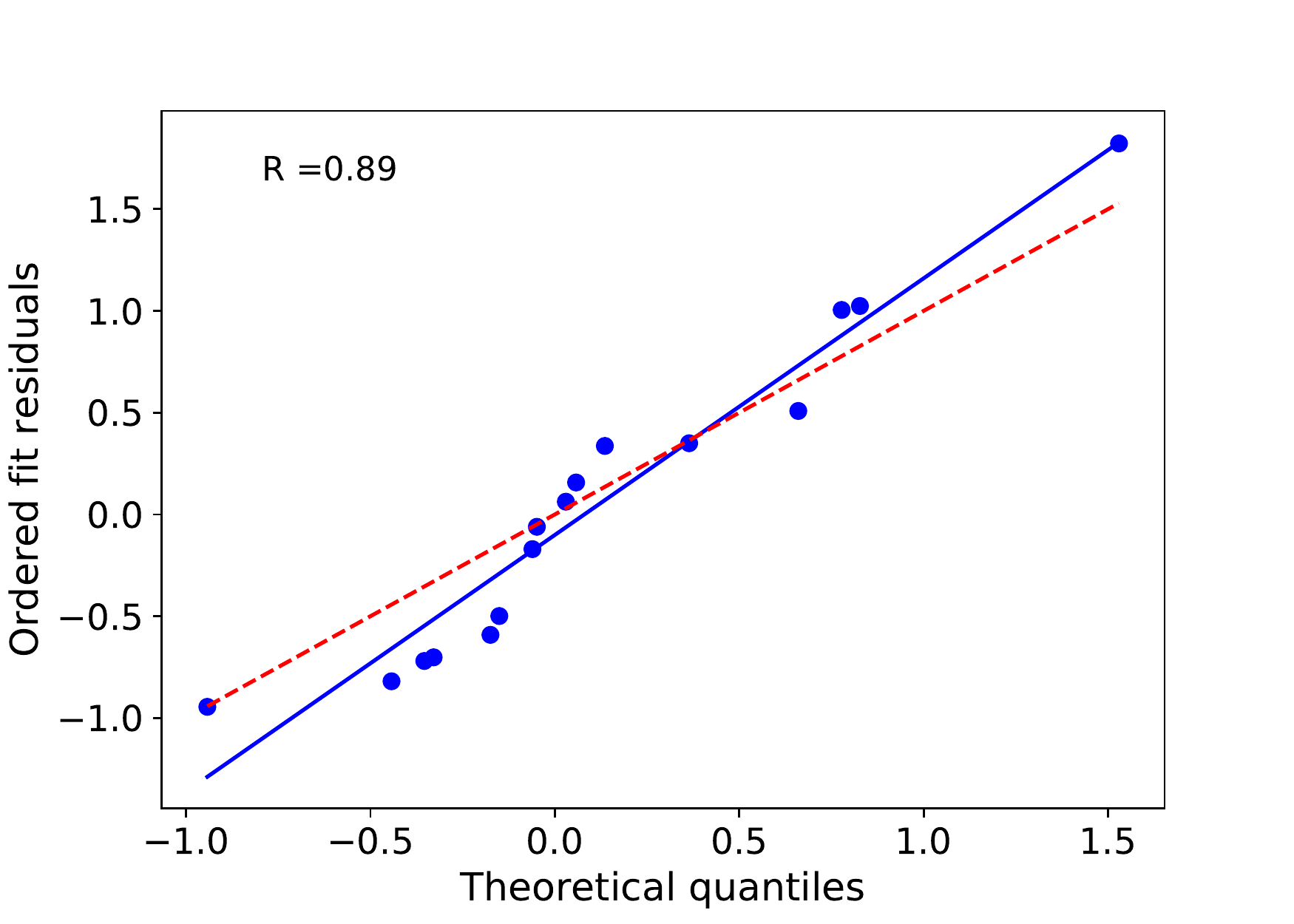}
\includegraphics[scale=0.5]{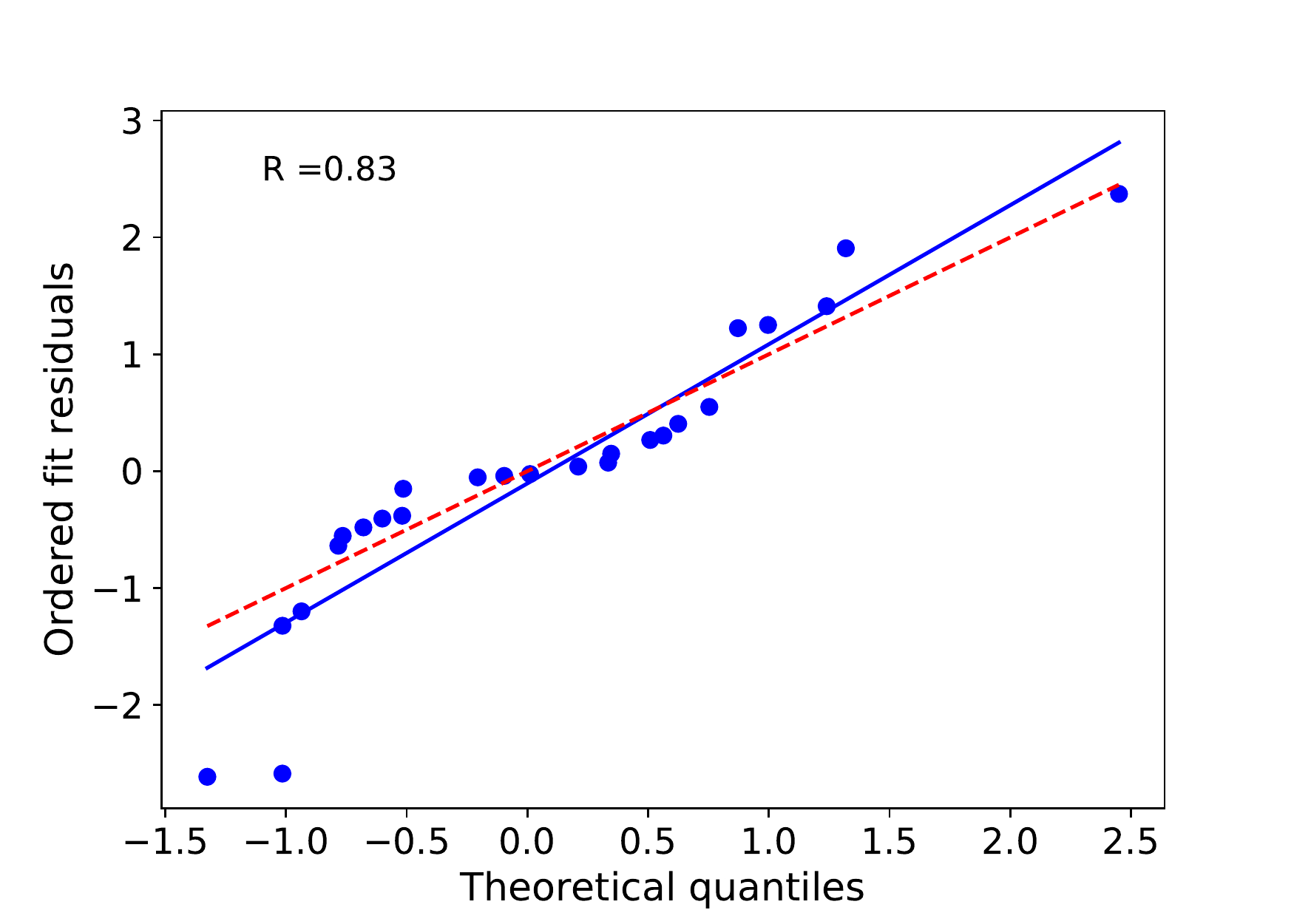}
\caption{Q--Q plot for the analysis of the Landau mass in SNM (top row) and energy per particle in PNM (bottom row). The reported $R$ score, see Eq.~\eqref{R_score}, is a measure of how well the blue points lie on the dashed red line.}
\label{fig:GF}
\end{figure}

In this section we quantify the goodness of the fits using the Q--Q plot method. We consider the two illustrative fits in Fig.~\ref{fig:GF}: the fit of the Landau mass in SNM using a quadratic density functional (top panel in Fig.~\ref{fig:ms:fit}) and the energy per particle in PNM as a function of density (lower right panel in Fig.~\ref{fig:scalings}).

For the analysis of the Landau mass, we found $\chi^2/\mathrm{dof} \approx 1.1$, and for the energy per particle, we find that $\chi^2/\mathrm{dof} \approx 1.4$. These numbers are already indications for a successful fit.

We further investigate the goodness of the fit by examining the normalized residuals of the fit, which are defined as the differences between the data and the best fit model.
These differences are expressed in the basis formed by the eigenvectors of the correlation matrix~\eqref{eq:correlation}. Then, each component of the eigenvectors are divided by the square root of the corresponding eigenvalue.
The assumption of this analysis for a good fit is that the normalized residuals should be uncorrelated and randomly distributed  around the mean value following a normal distribution.

Q--Q plots are a way for testing that the residuals follow a normal distribution. They are obtained as follow. First, the residuals are ordered from the smallest to the largest on the $y$-axis, then they are plotted against an ordered list of samples drawn from a normal distribution (centered at zero, with a width of one). If the residuals are perfectly normal distributed, then the result aligns on a straight line with slope $1$. This alignment can be captured by the coefficient of determination, $R$ defined as 
\begin{equation}
R = 1 - \frac{\sum_{i=1}^{n} (r_i - \hat{r}_i)^2}{\sum_{i=1}^{n} (r_i - \bar{r})^2}
\label{R_score}
\end{equation}
where $n$ is the number of residuals, $r_i$ are the residuals, $\hat{r}_i$ are the values expected for those residuals and $\bar{r} = \frac{1}{n} \sum_{i=1}^{n} r_i$. The best possible $R$ score is $1$ and it can be negative for arbitrarily worse fits.

The top panel of Fig.~\ref{fig:GF} shows the Q--Q plot for our analysis of the Landau mass in SNM. The ideal straight line with unit slope is shown as the dashed-red line, while the ordered residuals are shown as blue circles. The blue solid line is a fit to the blue points. The $R$ score, see Eq.~\eqref{R_score}, is $0.89$, which leads us to the conclusion that the ordered fit residuals are consistent with a normal distribution with a mean one, so close to $\chi^2/\mathrm{dof}=1$.
We also give similar results for the analysis of the energy per particle in PNM in the bottom panel of Fig.~\ref{fig:GF}. Here, the coefficient of determination is $0.83$, indicating a good fit.



\bibliography{biblio}

\end{document}


\title{Constraints on the nuclear symmetry energy from asymmetric-matter calculations with chiral NN and 3N interactions (Supplementary Material)}

\author{R. Somasundaram}
\email{somasundaram@ip2i.in2p3.fr}
\affiliation{Univ Lyon, Univ Claude Bernard Lyon 1, CNRS/IN2P3, IP2I Lyon, UMR 5822, F-69622, Villeurbanne, France}
 
\author{C. Drischler}
\email{cdrischler@berkeley.edu}
\affiliation{Department of Physics, University of California, Berkeley, CA 94720, USA}
\affiliation{Nuclear Science Division, Lawrence Berkeley National Laboratory,
 Berkeley, CA 94720, USA}

\author{I. Tews}
\email{itews@lanl.gov}
\affiliation{Theoretical Division, Los Alamos National Laboratory, Los Alamos, New Mexico 87545, USA}

\author{J. Margueron}
\email{j.margueron@ip2i.in2p3.fr}
\affiliation{Univ Lyon, Univ Claude Bernard Lyon 1, CNRS/IN2P3, IP2I Lyon, UMR 5822, F-69622, Villeurbanne, France}

\date{\today}

\begin{abstract}
We briefly describe the GitHub repository where the data and Python codes used in this paper are provided under the MIT license.
\end{abstract}

\maketitle

\section{Description of the GitHub repository}

The repository~\cite{rahul_2020_4010747} contains the Python codes used to perform the analysis as well as generate all the figures presented in the publication. This repository is publicly available on GitHub~\cite{rahul_2020_4010747} and distributed under the MIT license.

The model parameters are obtained from a Bayesian analysis using the nonlinear least-square fitting \mbox{\lsqfit}~package developed in Python~\cite{Lepage_lsqfit}, which uses the \textsc{scipy} least-squares minimization routine to optimize our model with respect to data.

\section{List of Folders}

There are two main directories at the root of the repository: data and results.

\begin{itemize}
\item data: contains the predictions (that we treat as data) from each of the six Hamiltonians considered in this work. They are given as ASCII files.

\begin{itemize}
\item data/Effective\_mass: consists of the single particle energies for the six Hamiltonians in symmetric and neutron matter. 
\item data/EOS\_Drischler: contains the data files for the energy per particle for different values of iso-spin asymmetry. 

\end{itemize}
\item results: contains the figures as pdf files. The following sub-folders contain the various figures presented in the paper as follows,

\begin{itemize}
\item results/3\_scales: Figs.~1 and~6.
\item results/crust\_core: Fig.~11.
\item results/effective\_mass: Figs.~2,~3,~4 and~5.
\item results/esym\_esym2: Figs.~7 and~8.
\item results/non\_quadraticities: Figs.~9 and~10.
\item results/goodness: plots presented in the appendix of the paper.
\end{itemize}

\end{itemize}

\section{List of Python codes}
\label{sec:script}

The Python codes are located in the main folder.
They depend on the following Python libraries: scipy, numpy, matplotlib, gvar, lsqfit~\cite{Lepage_lsqfit}, sklearn.metrics.

\begin{itemize}
\item EM.py: performs the analysis of the single particle energies and produces the plots stored in the folder results/effective\_mass.
\item SM\_NM.py: performs the analysis of the energy per particle in symmetric and neutron matter and produces the plots stored in the folder results/3\_scales.
\item symmetry\_energy.py: calculates the global symmetry energy and creates the corresponding plot stored in the folder results/esym\_esym2.
\item quadratic\_symmetry\_energy.py: calculates the quadratic symmetry energy and creates the corresponding plot stored in the folder results/esym\_esym2.
\item non\_quadraticities.py calculates the non-quadratic contributions to the symmetry energy as well as the final fit residuals and creates the plots stored in the folder results/non\_quadraticities.
\item crust\_core.py calculates the crust-core transition and produces the plot in the folder results/crust\_core.
\end{itemize}

\section{Launch the Python script}

The scripts are written in Python~3, and can be launched from a terminal as:
\begin{verbatim}
   python3 ScriptName.py
\end{verbatim}
where ``ScriptName.py'' is one of the scripts listed in Sec.~\ref{sec:script}.

\bibliography{supp}